\documentclass[12pt]{article}

\RequirePackage{amsthm,amssymb,amsfonts,graphicx,subfigure,caption,bm,float, fullpage,wrapfig,setspace}
\RequirePackage[colorlinks,citecolor=blue,urlcolor=blue]{hyperref}
\usepackage{lineno}
\usepackage{amsmath,amssymb,hyperref}

\numberwithin{equation}{section}
\theoremstyle{plain}
                          
\theoremstyle{remark}         
\theoremstyle{definition} 

\newcommand\om{\Omega}
\newcommand\real{\mathbb{R}}

\newcommand\A{\mathcal{A}}

\newcommand\T{\mathcal{T}}
\renewcommand\S{\mathcal{S}}
\newcommand\bJ{\bm{J}}
\newcommand\bS{\bm{S}}

\newcommand\bx{\bm{x}}

\newcommand\blam{\bm\lambda}

\newcommand\bi{\begin{itemize}}
\newcommand\ei{\end{itemize}}

\newcommand{\ind}{\stackrel{\mathrm{ind}}{\sim}}

\def\I{{\mathbf{1}}}

\def\woMR#1{\w@MR#1MR#1MR\relax}%
\def\w@MR#1MR#2MR#3\relax{#2}

\def\@MR#1 #2\relax#3{%
 \href{http://www.ams.org/mathscinet-getitem?mr=#1}%
 {\MRfixed{#3}}}%

\def\MRfixed{MR\woMR}%

\usepackage{natbib}

\labelformat{section}{Section~#1}
\labelformat{figure}{Figure~#1}
\labelformat{table}{Table~#1}	

\title{A partial likelihood approach to tree-based density modeling and its application in Bayesian inference}
\author{Li Ma\thanks{Email: li.ma@duke.edu}\\ Department of Statistical Science\\ Duke University \and Benedetta Bruni\thanks{Email: benedetta.bruni@duke.edu}\\ Department of Statistical Science\\ Duke University}

\begin{document}

\maketitle

\begin{abstract}
Tree-based priors for probability distributions are usually specified using a predetermined, data-independent collection of candidate recursive partitions of the sample space. To characterize an unknown target density in detail over the entire sample space, candidate partitions must have the capacity to expand deeply into all areas of the sample space with potential non-zero sampling probability. Such an expansive system of partitions often incurs prohibitive computational costs and makes inference prone to overfitting, especially in regions with little probability mass. Thus, existing models typically make a compromise and rely on relatively shallow trees. This hampers one of the most desirable features of trees, their ability to characterize local features, and results in reduced statistical efficiency. Traditional wisdom suggests that this compromise is inevitable to ensure coherent likelihood-based reasoning in Bayesian inference, as a data-dependent partition system that allows deeper expansion only in regions with more observations would induce double dipping of the data. We propose a simple strategy to restore coherency while allowing the candidate partitions to be data-dependent, using Cox’s partial likelihood. Our partial likelihood approach is broadly applicable to existing likelihood-based methods and, in particular, to Bayesian inference on tree-based models. We give examples in density estimation in which the partial likelihood is endowed with existing priors on tree-based models and compare with the standard, full-likelihood approach. The results show substantial gains in estimation accuracy and computational efficiency from adopting the partial likelihood.
\end{abstract}

\noindent%
\vfill

\newpage

\section{Introduction}
\label{sec:intro}

Bayesian modeling for probability distributions based on trees typically involves specifying a prior on the space of recursive partitions. While allowing a rich collection of potential partition locations to be supported under such a prior is desirable for effectively characterizing the underlying distribution, incorporating flexible partition locations substantially expands the space of candidate trees, leading to challenges in effectively sampling over such spaces. This is a core challenge in efficient and scalable Bayesian tree-based inference, especially in multivariate and high-dimensional settings. Consequently, some  well-known tree-based generative models in the Bayesian literature have focused on trees with constrained partition locations, most commonly requiring the splits in the recursive axis-aligned partitions to occur at predetermined, data-agnostic locations along one of the dimensions (e.g., the midpoint). Some famous examples include the Pólya tree \citep{ferguson1973,lavine1992} and tail-free processes \citep{freedman1963} in nonparametric Bayesian modeling. 
These restricted partition schemes specified without regard to the actual data at hand often result in excessive partitioning in low-density regions with few observations, causing overfitting (or high variance) and unnecessary computation, while not dividing the sample space deep enough in regions with a high concentration of data, resulting in underfitting (or large bias) there. 

Some previous works attempt to address overfitting by incorporating regularization into the modeling. For example, the optional Pólya tree (OPT) \citep{wongandma2010,ma2017} adopts the ideas from Bayesian CART \citep{chipman1998} in incorporating randomized stopping, a form of hard thresholding to avoid excessive partitioning. This type of  regularization is achieved through the prior specification only {\em after} the sampling model (or the likelihood) has been specified, and therefore it can only address the overfitting arising from an overparameterization in the likelihood in parts of the sample space where the predetermined partition tree is  deep enough, but is incapable of addressing the underparameterization arising from insufficiently deep trees that cause underfitting on those parts of the sample space with high probability mass and rich distributional structure.  

Such issues are not unique to Bayesian approaches to tree learning. In fact, non-Bayesian tree-based algorithms usually consider partitions that divide on or midway in between the {\em observed} data values, which greatly reduces the complexity of the space of candidate partition trees. Moreover, by considering splits along marginal empirical quantiles (such as the median), they alleviate the local over- and underfitting issues by forcing all tree nodes at the same depth to contain roughly the same number of data points. See, for example, a recent algorithm introduced for density estimation \citep{waltherandzhao2023} that employs such a design. These approaches generally deviate from likelihood-based inference and sacrifice many benefits that come with using the likelihood \citep{berger1988likelihood}.

A natural question is then whether one could incorporate data-dependent partition locations into likelihood-based inference so that the sampling model and hence the likelihood is parameterized in a way that makes downstream inference more efficient and robust. In the absence of oracular knowledge about the actual population marginal quantiles, a straightforward if not naïve approach is to adopt marginal empirical quantiles as the split points in the partition tree. This, of course, runs the risk of ``double dipping'' the data. A simple remedy is  data-splitting---by dividing the training data into two parts, using one part to define partition locations and the other to estimate the underlying density based on the chosen partition grid. 
While valid, this approach can ``waste'' a significant portion of the data and cause a decay in the frequentist risk of the resulting estimator.

We exploit a new ``data splitting'' strategy that operates on the basis of a dyadic recoding of real-valued observations into a sequence of 0's and 1's and then performs data splitting on such sequences. Consequently, the basic unit of empirical evidence is no longer a single observation $x$ from the corresponding unknown distribution, but the individual 0s and 1s in the corresponding sequence for $x$. This allows a smaller fraction of the data to be used to pin down the partition locations. Intuitively, if $x$ is only used to guide the choice of partition grids within a subset $A$ of the sample space, then the event that the data point arises in $A$, that is, $x\in A$, should still contribute to the learning of the underlying density. The naïve data-splitting strategy would discard $x$ in its entirety, but the dyadic recoding will preserve such information. 

While the dependency on the same data in specifying the partition tree as those used to estimate the underlying distribution appears to be at odds with pure likelihood-based inference, we show that through the dyadic recoding such a strategy is in essence  carrying out inference based on Cox's partial likelihood \citep{cox1975}. A practical consequence is that recipes for Bayesian inference under popular priors on tree-based density models can proceed based on the partial likelihood with only minor modifications. We use the OPT prior \citep{wongandma2010} to demonstrate how such inference can proceed, and show empirically the improvement in estimation accuracy that arises from the more effective data-dependent parametrization of the underlying density.

\section{Method}

\subsection{Likelihood decomposition along a dyadic partition tree}
We start by presenting a dyadic recoding of data that leads to equivalent likelihood functions under i.i.d.\ sampling. Suppose we observe training data $x_1,x_2,\ldots,x_n\in \om\subset \real^{d}$ that are i.i.d.\ samples from an unknown distribution $F$ with density $f=dF/d\mu$. 
Without loss of generality, assume that $\om$ is a compact rectangle, e.g., $[0,1]^{d}$. To begin, we assume $d=1$, but will consider the case $d>1$ starting in Section~\ref{sec:multivar}. 

The full likelihood of this model is given by $L(f;\bx) = \prod_{i=1}^{n} f(x_i)$. Let $\T$ be an {\em a priori} given dyadic partition tree over $\om$ that generates the Borel $\sigma$-algebra. $F$ can be parameterized by the binomial splitting probability at each tree split $F(A_l|A)$. For any node $A$ in $\T$, let $L(f;\bx,A)=\prod_{x\in A} f(x|A)$ be the conditional likelihood on the subtree rooted at $A$. In particular, $L(f;\bx,A)=L(f;\bx)$ for $A=\Omega$. Then, for a node $A$ with $n(A)>1$, where $n(A)$ denotes the number of observations in $A$, one can write $L(f;\bx,A)$ recursively as $L(f;\bx,A) = m_{{\rm fix}}(A) \cdot L(f;\bx,A_l) L(f;\bx,A_r)$
where $A=A_l\cup A_r$ is the dyadic split of $A$ according to $\T$ and 
\begin{align}
\label{eq:decomp_fix}
m_{{\rm fix}}(A)=\prod_{x\in A} {\rm P}(x\in A_l|A)^{\I(x\in A_l)}\cdot {\rm P}(x\in A_r|A)^{\I(x\in A_r)}=F(A_l|A)^{n(A_l)} F(A_r|A)^{n(A_r)}.
\end{align}

\noindent If $n(A)\leq 1$, further decomposition is unnecessary. This is because for $n(A)=0$, $L(f;\bx,A)=1$, and for $n(A)=1$, $L(f;\bx,A) = f(x|A)$ where $x$ denotes the single observation in $A$. For continuous distributions, the probability that any data fall exactly on the boundary of any node in $\T$ is nil, and hence for simplicity, we do not explicitly describe this possibility.

This full likelihood has been the basis for several Bayesian priors on $f$ such as P\'olya trees and its various generalizations that endow priors on $F(A_l|A)$. Some recent developments including OPT further incorporate $\T$ into the model by specifying a prior on the recursive partition tree $\T$. However, this fully Bayesian approach requires that the prior on $\T$ ignores the observed data at hand. 

Now, let us consider a different decomposition of the full likelihood above along a dyadic tree that is {\em not} given {\em a priori} but rather with splitting points sitting right on top of an observed data point. We shall now use $\T(\bx)$ to denote the tree, which emphasizes its dependence on the data. Again, we provide an inductive description of the construction of $\T(\bx)$ along with a likelihood decomposition based on $\T(\bx)$. Let $A\in \T(\bx)$ be a partition block or node that arises in our inductive procedure. We let $x_{A}\in A$ be an observed data point based on which we further divide $A$, and let $A=\{x_{A}\}\cup A_l\cup A_r$ where $A_l=\{x\in A: x<x_{A}\}$ and $A_r=\{x\in A: x>x_{A}\}$ are the two child nodes in $\T(\bx)$. Note that the definitions of $A_l$ and $A_r$ are different from before, but we shall use the same notation without causing confusion. Then we can write the likelihood recursively as follows
\begin{align}
\label{eq:decomp_new}
L(f;\bx,A) = m(A) \cdot L(f;\bx,A_l) \cdot L(f;\bx,A_r)
\end{align}
\noindent where $m(A)=f(x_{A}|A) \cdot F(A_l|A)^{n(A_l)} F(A_r|A)^{n(A_r)}$, and $L(f;\bx,A)$ is the conditional likelihood on the subtree rooted at $A$ given that $A$ arises as a node in the data-dependent tree.

Contrasting this new decomposition with what we have in Eq.~\eqref{eq:decomp_fix}, the observation $x_{A}$ is now {\em singled out} on $A$ and used to determine the partition of $A$ and is correspondingly eliminated from further recursive expression of $L(f;\bx,A_l)$ and $L(f;\bx,A_r)$. While this factorization holds for any observed value of $x_{A}$ in $A$, in practice it is often desirable to create partition trees that are evenly grown---that is, nodes in each level of the tree have almost the same number of observations. In that case one can choose $x_{A}$ to be the observed median in $A$. That is, let $x_{A}$ be the $k(A)$th order statistic on $A$ with $k(A)=\lfloor n(A) + 1\rfloor/2$. This will be adopted in all of our numerical examples.

Continuing this iteration, we can write the full likelihood in contributions from the tree nodes along a data-dependent tree $\T(\bx)$ as $L(f;\bx,\om)=\prod_{A\in\T(\bx)} m(A)$.
More generally, if we let $\T_{A}(\bx)$ be the subtree of $\T(\bx)$ rooted at $A$, then $ L(f;\bx,A)=\prod_{A'\in \T_A(\bx)} m(A')$. Here we note that for any dataset with finite sample size, there will be only a finite number of non-empty nodes in the data-dependent tree. So $\T(\bx)$ is always a finite tree.

\subsection{A partial likelihood approach to inference}
\label{sec:partial_like}
The data-dependent nature of the above full likelihood decomposition hampers fully likelihood-based strategies such as Bayesian inference. At the same time, we recognize a striking similarity of the likelihood contribution on node $A$ in a data-dependent tree, $m(A)$, compared to the corresponding likelihood contribution $m_{{\rm fix}}(A)$ with a data-independent fixed tree. Both have a binomial likelihood component corresponding to those points that fall into $A_l$ versus $A_r$. The only difference is the extra component $f(x_A|A)$ in $m(A)$. 
However, we make two observations: (i) the branching probabilities $F(A_l|A)$ and $F(A_r|A)$ are sufficient for determining the probability distribution $F$ almost everywhere on $\om$; (ii) the density at the exact partition boundaries cannot be reliably inferred anyway, even with a fixed partition tree, as those are usually the boundaries of piecewise constant densities under tree-based models.

Consequently, we proceed by treating the exact conditional density of $f(\cdot|A)$ at the point $x_{A}$, i.e., $f(x_A|A)$, as ``nuisance'' parameters. Following the seminal work on the proportional hazards model \citep{cox1972}, \cite{cox1975} formulated the notion of partial likelihood more generally by rewriting a data set $\bx$ as $\bx=(w_1,\bx_1,w_2,\bx_2,\ldots,w_N,\bx_N)$ and assumed that the full likelihood can be written as $L(f;\bx)=\prod_{j=1}^{N}f(w_j|d_j) \prod_{j=1}^{N} f(\bx_j|c_j)$ where $d_j=(w_1,\bx_1,\ldots,w_{j-1},\bx_{j-1})$ and $c_j=(w_1,\bx_1,\ldots,w_{j-1},\bx_{j-1},w_{j})$. The first product generally involves some nuisance parameters. \cite{cox1975} refers to the second product in the factorization, $\prod_{j=1}^{N} f(\bx_j|c_j)$, as the ``partial likelihood''. 

To see that under the data-dependent tree, we have a likelihood factorization exactly of this form, one can simply replace the index $j$ in Cox's notation with a coarse-to-fine indexing of the nodes in a dyadic tree. There are various ways to define such an index. All we need is to ensure that it is consistent with a partial order of the nodes, so that a parent node always has a lower index $j$ than its children. For example, we can simply let $j$ be the order of a node $A$ corresponding to the dyadic sequence of 0 and 1 corresponding to whether a descendant of $A$'s is the left or right child of a split. Now $w_j$ corresponds to $x_A$, the value on which we divide the node $A$. In this case, $f(w_j|d_j)$ is  
$f(x_A|A)$. On the other hand, each $\bx_j$ corresponds simply to the collection of indicators of whether each point in $A$ falls into $A_l$ or $A_r$, whose conditional likelihood given $c_j$ depends only on the parameters of interest, here the branching probabilities $F(A_l|A)$ and $F(A_r|A)$. 

We can write the tree-based likelihood decomposition in Cox's form. 
Specifically, 

\[
L(f;\bx,\om) = \prod_{A\in\T(\bx)} f(x_{A}|A) \cdot \prod_{A\in\T(\bx)}m_{\rm P}(A) 
\]

\noindent where 
$m_{\rm P}(A)=F(A_l|A)^{n(A_l)}F(A_r|A)^{n(A_r)}$. The second piece in this product gives Cox's partial likelihood $ L_{{\rm P}}(f;\bx,\om) = \prod_{A\in \T(\bx)} m_{{\rm P}}(A),$ which can be written recursively as
\begin{align}
\label{eq:decomp_partial}
L_{{\rm P}}(f;\bx,A) = m_{{\rm P}}(A) \cdot L_{{\rm P}}(f;\bx,A_l) \cdot L_{{\rm P}}(f;\bx,A_r)= \prod_{A'\in\T_{A}(\bx)} m_{{\rm P}}(A').
\end{align}

It is remarkable that this partial likelihood has essentially the same form as the full likelihood under a fixed-tree decomposition. This implies that inference strategies based on the full likelihood on a fixed tree can be applied with little change to this partial likelihood by acting {\em as if} the data-dependent tree were actually fixed.

One question regards how much information is discarded by ignoring the ``nuisance'' portion of the likelihood, namely $\prod_{A\in\T(\bx)} f(x_{A}|A)$. 
It is worth noting that the information discarded is much less than a conditional likelihood approach such as the one proposed in \cite{lavine1995}, which conditions on the full set of observed partition points. To see this, note that the total  contribution from the partition points to the full likelihood is $\prod_{A\in\T(\bx)} f(x_{A}),$
where $f(x_{A})$ is the density of $f$ (not conditional on $x_{A}$ being in $A$) evaluated at $x_{A}$. That is, $f(x_{A})= f(x_{A}|A)\cdot F(A).$
In other words, the partial likelihood retains the information about $F$ from the event that the corresponding partition point on $A$ is an observation {\em in} $A$, and only discards the information provided by its {\em relative value within $A$}. Another way to see this is that the partition point $x_{A}$ contributes to the binomial likelihood on ancestral nodes of $A$ in the partial likelihood because

\[
F(A)=\prod_{A'\in\A(A)}F(A'_l|A')^{\I(x_{A}\in A'_l)}F(A'_r|A')^{\I(x_{A}\in A'_r)}
\]

\noindent where $\A(A)=\{A'\in\T(\bx): A\subsetneq A'\}$ are the ancestors of $A$ in $\T(\bx)$. The binomial likelihood terms on all ancestors of $A$ are indeed retained in the partial likelihood. For this reason, the partial likelihood approach is in essence carrying out data splitting using the 0 and 1 dyadic sequence---preserving the prefix of the sequence that indicates $x\in A$, and discarding the postfix that encodes the relative value of $x$ in $A$. 

It is worth noting that the partial likelihood approach, with a slight generalization, provides a unifying framework that contains the full likelihood approach based on pre-determined partition locations as a special case and allows the approach to be applied to distributions that might not be continuous. To see this, in the decomposition in Eq.~\eqref{eq:decomp_new}, we can let $m(A)=f(x_{A}|A)^{n(\{x_A\})} \cdot F(A_l|A)^{n(A_l)} F(A_r|A)^{n(A_r)}$ where $n(\{x_A\})$ is the number of observations exactly equal to $x_A$. In the previous situations, the fixed and observed partition locations correspond to the special case when $n(\{x_A\})=0$ and $n(\{x_A\})=1$, respectively. Under this unified framework, one can use a mix of partition locations that include some that sit on one or more observations and others not on any observations. Inference can then proceed using the partial likelihood as defined in Eq.~\eqref{eq:decomp_partial} in the same way without changes.

\subsection{Partial likelihood for multivariate tree-based density models}
\label{sec:multivar}

The discussion so far has concerned a univariate sample space $[0,1]$ on which recursive partitions are defined. Many problems in which tree-based methods are useful involve multivariate sample spaces, e.g., $[0,1]^d$, in which there are multiple axis-aligned partition directions that can be chosen to grow the dyadic partition tree. The partition tree can then be used to specify models on $f$ and the likelihood can serve as the basis for inferring what the underlying partition tree should be to generate models on $f$ that fit the data well. 

One principled strategy to incorporating inference on the tree topology is to treat the splitting variables, denoted by $\bJ$, which determine the tree topology $\T(\bJ)$ as latent variables. In this way, the full likelihood is expressed as a likelihood of $f$ and $\bJ$ jointly. In particular, we utilize the following decomposition
\[
L(f,\bJ;\bx,A) = 
m_{{\rm fix},J(A)}(A) \cdot L(f,\bJ;\bx,A_{J(A),l})\cdot L(f,\bJ;\bx,A_{J(A),r})
\]

\noindent where for $j\in \{1,2,\ldots,d\}$, $A_{j,l}$ and $A_{j,r}$ denote the two children nodes if $A$ is divided in the $j$th direction at a pre-determined fixed location and $m_{{\rm fix},j}(A)=F(A_{j,l}|A)^{n(A_{j,l})} F(A_{j,r}|A)^{n(A_{j,r})}.$
Treating the splitting variables as latent auxiliary parameters, one can use this decomposition to infer jointly the density $f$ and the latent variables $\bJ$ \citep{wongandma2010}.

One can naturally apply the same strategy to the unified partial likelihood framework regardless of whether or not partition locations sit on observed values. The only complication is that now the tree topology depends on both $\bJ$ and the observed data $\bx$. Thus we denote the tree by $\T(\bJ,\bx)$. The full likelihood for $(f,\bJ)$ can then be decomposed as follows:
\begin{align*}
L(f,\bJ;\bx,A) &=m_{J(A)}(A) \cdot L(f,\bJ;\bx,A_{J(A),l})\cdot L(f,\bJ;\bx,A_{J(A),r})= \prod_{A'\in\T_{A}(\bJ,\bx)} m_{J(A')}(A')
\end{align*}
for each $A\in\T(\bJ,\bx)$, where $A_{j,l}=\{\bx\in A:x_j<x_{A,j}\}$, $A_{j,r}=\{\bx\in A:x_j>x_{A,j}\}$ and $m_j(A) = f(x_{A,j}|A) F(A_{j,l}|A)^{n(A_{j,l})} F(A_{j,r}|A)^{n(A_{j,r})}$ for $j\in \{1,2,\ldots,d\}$.

It is important for the partition location on each dimension $j$ to be through the same observation $x_A$ to ensure the validity of the partial likelihood. This is because otherwise the nuisance parameter $f(x_A|A)$ would be different for different $j$, resulting in a different piece of the likelihood to be discarded for different ways of partitioning. Such discrepancies could accumulate over multiple levels of partitions, rendering the partial likelihood corresponding to different partition sequences incomparable, and thus unfit for learning the partition tree. As such, one would need an effective way to choose the partition point $x_A$. The ideal $x_A$ should be close to the marginal median along each dimension of $A$. Searching for such $x_A$'s might be computationally challenging depending on the nature of the underlying distribution~$F$. 

An alternative strategy is to simply pick an observation $x_A^j$ to divide on specifically for each dimension $j$, and then take out all of these potential division points $\bx_A=\{x_A^j:j=1,2,\ldots,d\}$. The nuisance part of the likelihood under each partition direction $j$ then becomes equal, and for each dimension $j$, $m_j(A) = \left(\prod_{j=1}^{d} f(x_A^j|A)\right) \cdot F(A_{j,l}|A)^{n(A_{j,l})} F(A_{j,r}|A)^{n(A_{j,r})}$,  but now $A_{j,l}=\{\bx\in A:x_j<x^j_{A,j}\}$, $A_{j,r}=\{\bx\in A:x_j>x^j_{A,j}\}$ and $n(A_{j,l})$ and $n(A_{j,l})$ are the number of observations in $A_{j,l}$ {\em excluding} all those in $\bx_A$. Compared to dividing on the same $x_A$ in all directions, this alternative strategy discards $d$ times as much data, but avoids the difficulty in choosing the single $x_A$ as now one can again simply divide marginally on, for example, the median. This trade-off is often favorable when $d$ is small (e.g., $\leq 5$).

Under either strategy, similar to before, we can then define the partial likelihood as

\begin{align}
\label{eq:likelihood_multi_partial}
L_{{\rm P}}(f,\bJ;\bx,A)=\prod_{A\in\T_A(\bJ,\bx)} m_{{\rm P},J(A')}(A')
\end{align}

\noindent where for $j\in\{1,2,\ldots,d\}$, $ m_{{\rm P},j}(A)=F(A_{j,l}|A)^{n(A_{j,l})} F(A_{j,r}|A)^{n(A_{j,r})}.$
The partial likelihood can then be used to infer jointly $(f,\bJ)$ as with the full likelihood. We will demonstrate later through numerical experiments in Section~\ref{sec:num_ex_bivariate}.

\section{Bayesian modeling based on the partial likelihood}
\label{sec:bayes}

Given the striking similarity between $m_P(A)$ and $m_{{\rm fix}}(A)$, it is not surprising then that existing tree-based Bayesian models for $F$ and $f$, which are constructed through specifying priors on the splitting probabilities $F(A_l|A)$, can apply readily to the case with data-dependent tree splits with the partial likelihood. We shall consider a few of these examples. Before presenting these examples, quickly review a technique that expresses the likelihood of $f$ with respect to a baseline density $h$, which is not necessary but can substantially simplify our notation and derivation for the posterior distribution in the examples.

\subsection{Expressing the partial likelihood in terms of likelihood ratios}

One technique introduced in \cite{ma2017} that simplifies the expression of the likelihood under a fixed tree partition and leads to more numerical robustness is to express the full likelihood decomposition with respect to that under a base measure. This technique can be readily applied to the current partial likelihood. Specifically, let $H$ be a fixed baseline measure on $\om$, and let $h=dH/d\mu$ be its density. Then the partial likelihood under this baseline density $L_{\rm P}(h;\bx,\om)$. We can then write the partial likelihood ratio between any model $f$ and $h$:
\[ 
L_{\rm P}(f;\bx,\om)/L_{\rm P}(h;\bx,\om)=\prod_{A\in \T(\bx)} r_{\rm P}(A) \qquad \text{where}\quad  
r_{\rm P}(A)=\frac{F(A_l|A)^{n(A_l)} F(A_r|A)^{n(A_r)}}{H(A_l|A)^{n(A_l)} H(A_r|A)^{n(A_r)}}.
\]

That is, the overall (partial) likelihood ratio between $F$ and $H$ can be factorized over the tree nodes where $r_P(A)$ is the (partial) likelihood ratio with respect to the corresponding binomial likelihood. Accordingly, we can express this factorization in each subtree $\T_{A}(\bx)$ as $L_{\rm P}(f;\bx,A)/L_{\rm P}(h;\bx,A)=\prod_{A'\in \T_{A}(\bx)} r_{\rm P}(A').$

The benefits of expressing the (partial) likelihood ratio with respect to that of a base model stem mainly from the fact that likelihood ratios are invariant with respect to the change of variable from the original observation $\bx$ to some transformed observations $\bx'$. 
Accordingly, the likelihood ratio, full or partial, is then scale-free with respect to any change of scales on the measurement of $\bx$. This scale-invariant property will become especially handy in simplifying the computation of inference algorithms and avoiding numerical problems.

\subsection{Example 1: P\'olya trees}
\label{sec:pt}

The first example we consider is the P\'olya tree (PT) model. PT is the simplest model on $F$ based on prior specifications on $F(A_l|A)$. It assumes simple independent beta priors on $F(A_l|A)$, which renders simple conjugate posterior inference. Now we can again adopt this prior, but now carry out (partial) Bayesian inference through a conjugate posterior update with the partial likelihood. An important simplification of the current framework is that the data-adaptive tree will {\em always} be finite for any finite sample as eventually all data points will be partitioned upon. For this reason, we only need to consider the so-called finite PT, which are PT priors defined on trees of a finite number of levels, and the model will {\em always} generate a density function.

Specifically, PT models the branching probabilities independently on all interior nodes in the partition tree using beta priors $F(A_l|A) \ind {\rm Beta}(\alpha_l(A),\alpha_r(A)).$
On a leaf node $A$, 
simply set $F(\cdot|A)=u_A(\cdot)$ to some base distribution $u_A$ supported on~$A$. A popular way to specify the beta priors is by using a concentration (or shrinkage) parameter along with a mean base measure. Specifically, we can let $H$ be a probability measure on $\om$. Then if we let $\alpha_l(A)=c(A) H(A_l|A)$ and $\alpha_r(A)=c(A) H(A_r|A)$ where $c(A)$ is possibly the node-specific concentration parameter on $A$. If in addition, we let the base distribution on a leaf node $A$ be $u_A(\cdot)=H(\cdot|A)$, then the resulting PT prior will have its prior mean equal to $H$. 
Bayesian inference (using the partial likelihood) can proceed now due to beta-binomial conjugacy, and the posterior model is simply a PT given the same data-adaptive tree with $F(A_l|A)\,|\,\bx \ind {\rm Beta}(\alpha_l(A)+n(A_l),\alpha_r(A)+n(A_r)).$

One could also compute the overall marginal likelihood after integrating out the PT prior. Specifically, due to the factorization of the partial likelihood over the tree nodes and the prior independence of $F(A_l|A)$ for all $A$s, the ``partial'' marginal likelihood is given by $\Phi(\om)=\int L_{\rm P}(f;\bx,\om) \pi({\rm d}F)=\prod_{A\in\T} M_{{\rm P}}(A)$ where 

\[
M_{{\rm P}}(A):=\int m_{\rm P}(A)  \pi(d F(A_l|A))=\frac{B(\alpha_l(A)+k(A)-1,\alpha_r(A)+n(A)-k(A))}{B(\alpha_l(A),\alpha_r(A))}\]

\noindent with $B(a,b)$ being the beta function. That is, $B(a,b)=(\Gamma(a)\Gamma(b))/\Gamma(a+b)$ for $a,b>0$.

The posterior predictive density---that is, the posterior mean of $f$---is also available exactly. To evaluate the posterior predictive density at a point $\bx' \in \om$, we simply need to trace the sequence of nodes in $\T$ given by the training data, $\bx'\in A_{k}\subset A_{k-1}\subset\cdots \subset A_{0}=\om$ where $A_k$ is a leaf, that is, $n(A_k)\leq 1$. Then the posterior mean density at $\bx'$ is
\[
{\rm E}(f(\bx')|\bx) = {\rm E}(F(A_k)|\bx)\cdot h(\bx'|A_k)=\left(\prod_{i=0}^{k-1} {\rm E}\left(F(A_{i+1}|A_{i})|\bx\right)\right)\cdot h(\bx'|A_k)
\]
where ${\rm E}\left(F(A_{i+1}|A_{i})|\bx\right)$ is the corresponding posterior mean of $F(A_{i+1}|A_{i})$. That is
\begin{align*}
{\rm E}\left(F(A_{i+1}|A_{i})|\bx\right)
&= \begin{cases}
\frac{\alpha_l(A_{i})+n(A_{i+1})}{\alpha_l(A_{i})+\alpha_{r}(A_{i)}+n(A_{i})-1} & \text{if $A_{i+1}$ is the left child of $A_i$}\\
\frac{\alpha_r(A_{i})+n(A_{i+1})}{\alpha_l(A_{i})+\alpha_{r}(A_{i})+n(A_{i})-1} & \text{if $A_{i+1}$ is the right child of $A_i$.}
 \end{cases}
\end{align*}
This predictive density can be expressed in a scale-free manner w.r.t.\ the base measure

\[
{\rm E}(f(\bx')|\bx) = h(\bx')\cdot \frac{{\rm E}(F(A_k)|\bx)}{H(A_k)} =h(\bx')\cdot \prod_{i=0}^{k-1} \frac{{\rm E}\left(F(A_{i+1}|A_{i})|\bx\right)}{H(A_{i+1}|A_{i})}.
\]

\noindent The posterior predictive density ratio w.r.t.\ the base density can also be computed recursively, 

\[
\frac{{\rm E}(f(\bx'|A_i)|\bx)}{h(\bx'|A_i)}=
\frac{{\rm E}\left(F(A_{i+1}|A_{i})|\bx\right)}{H(A_{i+1}|A_{i})}\cdot \frac{{\rm E}(f(\bx'|A_{i+1})|\bx)}{h(\bx'|A_{i+1})} 
\]

\noindent for $i=0,\ldots,k-1$, and it is equal to 1 for $i=k$. To emphasize the recursive nature, let $\xi(A)=\frac{{\rm E}(f(\bx'|A)|\bx)}{h(\bx'|A)}$ and let $\varphi(A)=\frac{{\rm E}\left(F(A|A_{p})|\bx\right)}{H(A|A_{p})}$. Then $\xi(A_k)=\varphi(A_k)=1$ and for $i=0,1,\ldots,k-1$,
$\xi(A_i) = \frac{{\rm E}\left(F(A_{i+1}|A_{i})|\bx\right)}{H(A_{i+1}|A_{i})}\cdot  \xi(A_{i+1})
=\varphi(A_{i+1})  \xi(A_{i+1})$,
which is a bottom-up recursion on the sequence $A_k,A_{k-1},\ldots,A_{0}$. While this recursive expression is not necessary for the PT model {\em per se}, as we will describe in the following subsection, when latent Markov state variables are incorporated into PT, such a recursive expression for the posterior predictive density ratio will become important.

In problems involving model choice and hypothesis testing \citep{bergerandguglielmi2001,maandwong2011, holmes2015, sorianoandma2017} it is often helpful to compute the likelihood ratio, that is, the Bayes factor between the PT model and the corresponding partial likelihood under the base density $h$, denoted by $\phi(\om)$, which is

\[
\phi(\om)=\frac{\int L_{\rm P}(f;\bx,\om) \pi({\rm d}F)}{L_{\rm P}(h;\bx,\om)}
=\prod_{A\in \T}\int r_{{\rm P}}(A)\pi({\rm d}F(A_l|A))=\prod_{A\in \T}\eta_{{\rm P}}(A),
\]

\noindent where $\eta_{{\rm P}}(A):=\int r_{{\rm P}}(A)\pi({\rm d}F(A_l|A))=\frac{M_{{\rm P}}(A)}{H(A_l|A)^{k(A)-1}H(A_r|A)^{n(A)-k(A)}}$, which is the ``evidence'' for $F$ marginalized over the prior with respect to the base measure $H$. We could also define the corresponding Bayes factor on the subtree $\T_A$ rooted at $A$, denoted by $\phi(A)$ to gauge the model against the base measure 
\[
\phi(A) = \prod_{A'\in\T_{A}}\eta_{{\rm P}}(A')=\begin{cases} \eta_{{\rm P}}(A)\phi(A_l)\phi(A_r) & \text{if $n(A)>1$}\\ 1 &\text{otherwise.}\end{cases}
\]

One benefit of the current approach is that we now know {\em a priori} the number of samples that fall into each tree node $A$ at the $l$th level of $\T$ before even observing the data, once the total sample size is known. The sample size $n(A)\approx n\cdot 2^{-(l(A)-1)}$. One can use this {\em prior knowledge} to specify the concentration parameter $c(A)$ depending on $l(A)$. 

It has been a popular strategy to specify $c(A)$ in terms of $l(A)$ in traditional PT modeling when the partition tree is not data-adaptive, as was recommended in \cite{lavine1992}. But \cite{ma2017} argued that for such models with predetermined fixed partition locations, one should generally {\em not} specify the concentration parameter based {\em solely} on the level of the tree, as it will not adapt to the proper level of smoothness in the underlying function. These papers introduced adaptive shrinkage prior in terms of putting priors on $c(A)$ in the form of a Markov process along the partition tree to resolve this issue.

However, with the location of the data-adaptive partition achieved with the partial likelihood, the simple strategy of specifying the concentration parameter $c(A)$ as a function of $l(A)$ appears less problematic. For example, when we always partition on the empirical median in each $A$, the number of samples that fall into every tree node $A$ in the $l$th level of $\T$ is always roughly $n\cdot 2^{-l}$. Now that the nodes in the same level of the tree contain roughly the same number of observations, the model complexity within each level of the tree is {\em automatically} re-scaled corresponding to the local concentration of mass in $F$. 

The predetermined sample size in the tree nodes brings an additional computational benefit. Because $n(A)$ is (roughly) always $n\cdot 2^{-l(A)}$, the beta posteriors within each tree level are essentially different only due to the prior beta pseudo-counts. If we adopt the same beta pseudo-counts in each level of the tree, then there are only a small collection of beta functions that need to be evaluated. If instead one adopts pseudo-counts based on a base measure such as $\alpha_l(A)=c H(A_l|A)$ and $\alpha_r(A)=c H(A_r|A)$, then the beta functions also depend on the relative sizes of $A_l$ and $A_r$ with respect to the base measure $H$, that is, $(H(A_l|A),H(A_r|A))$. Even in these cases, one may precompute them using a grid of values for these relative sizes, such as $(0.01,0.99),(0.02,0.98),\ldots,(0.99,1)$.
In applications such as hypothesis testing where the marginal likelihood is needed, the main computation involves evaluating the corresponding beta functions for the posterior and prior beta distributions for each $A$. Such an approximation can substantially speed up computation. This speed-up can be especially helpful in multivariate settings.

There is an interesting connection between this partial likelihood-based PT to a nonparametric process called quantile pyramids (QP) introduced by \cite{hjortandwalker2009}. QP is a prior distribution on the unobserved quantiles of $F$. It is designed to address issues caused by prefixed partition locations as well. It is computationally demanding to carry out inference with QP, as it breaks down the beta-binomial conjugacy in PT. Our partial likelihood approach provides an alternative that maintains the conceptual advantage of partitioning along quantiles (though now the empirical ones) while preserving the computational tractability of the PT. 

There exists an asymptotic relationship between our partial likelihood-based PT model and the QP. To see this, suppose now $n\rightarrow \infty$ and we choose $k(A)$ to be a fixed empirical $q$th quantile ($q\in (0,1)$) such as the median (i.e., $q=0.5$) in $A$, then under regularity conditions $\T(\bx)$ will converge with probability 1 to the dyadic tree defined by the dyadic quantiles of $\om$. Moreover, for each $A\in \T$, $n(A)\rightarrow \infty$ and $k(A)/n(A)\rightarrow q$. Thus, asymptotically, the partition tree $\T(\bx)$ will converge to the population quantile pyramid tree. 

Interestingly, \cite{hjortandwalker2009} proposed to address the computational intractability of the QP by adopting a multinomial approximate likelihood introduced in \cite{lavine1995}, which replaces the full likelihood under $f$ with a finite discrete approximation based on the quantile grid of $F$ up to a certain resolution. This approach leads to a posterior computational algorithm similar to the conjugate computation we achieve with the partial likelihood. 
However, it is acknowledged in  \cite{lavine1995} that such an approximation generally introduces bias and leads to overly conservative inference. Our partial likelihood approach avoids this issue and preserves the consistency of the inference.

\subsection{Example 2: Models that incorporate Markov latent states}
\label{sec:latent_state}

The collection of beta concentration parameters $\{c(A):A\in \T\}$ plays a critical role in regularizing the smoothness of $f$ relative to the base density $h$. Previous works \citep{wongandma2010,ma2017,christensenandma2020} have shown that inference can be substantially enhanced under PT by incorporating these concentration parameters into the modeling and learning them based on the data. They introduce a technique that allows adaptive inference on these concentration parameters to characterize potentially spatially heterogeneous features of $F$ while maintaining computational tractability using latent variables with a first-order Markov model. We show how these models can also be applied with partial likelihood.

Specifically, given a partition tree $\T$, we consider PT models specified conditional on a latent state variable $S(A)$ associated with the node $A$. In other words, the branching probabilities are modeled as a mixture of betas with $S(A)$ representing the corresponding mixture component to which $F(A_l|A)$ belongs. That is $F(A_l|A)\,|\,S(A)=s \ind {\rm Beta}(\alpha_{s,l}(A),\alpha_{s,r}(A))$
and jointly modeled the states using a root-to-leaf Markov process with a (prior) transition matrix $\rho_{s,s'}(A)$ on each node $A$ to move into $s'$ given $S(A_p)=s$. Two special cases are when $\rho_{s,s}(A)=1$ for all $A$ and $s$, corresponding to all states being identical, and when $\rho_{s,s'}(A)=\rho_{s'}(A)$ for all $A$, $s$ and $s'$, corresponding to all states being independent across~$A$'s. 

One can compute the posterior transition matrix of the Markov tree model exactly based on a one-pass message passing (essentially a forward-backward algorithm for discrete Markov process). More specifically, a key component of the message passing algorithm is the marginal likelihood of the submodel in $A$ given the parent state, that is, $S(A_p)=s$. To emphasize the recursive nature of this algorithm, we define this marginal likelihood on $A$ as $\Phi_{s}(A)$, which is given by the following recursive formula $\Phi_{s}(A)=\sum_{s'\in \S}\rho_{s,s'}(A)M_{s'}(A)\Phi_{s'}(A_l)\Phi_{s'}(A_r)$
where $M_{s}(A):=\int m(A) \cdot \pi_s({\rm d}F(A_l|A))$ for all $s$ and 
$\pi_s({\rm d}F(A_l|A)):=\pi({\rm d}F(A_l|A)|S(A)\!=\!s)$.

Based on a bottom-up recursion from the leaf to the root, one can compute these mappings for all $A\in \T(\bx)$ and $s\in\S$. Note that the above expression applies to any prior specification on $F(A_l|A)$. When beta priors are adopted under the PT, the integral boils down to the reciprocal of a beta function with a closed form. 

We can normalize the mapping $\phi_s(A)$ with respect to the likelihood of a base measure $H$. Specifically, let $h$ be the corresponding density of the base measure on $\om$ around which the PT is centered. We can define a marginal likelihood ratio mapping $\phi_s(A) = \Phi_s(A)/\prod_{x\in A} h(x|A).$
The likelihood ratio mapping can again be expressed recursively. For any $A$ with $n(A)>1$,
\begin{align*}
\phi_{s}(A)
&=\sum_{s'\in \S}\rho_{s,s'}(A)\eta_{s'}(A)\phi_{s'}(A_l)\phi_{s'}(A_r)
\end{align*}
where $r(A)=\frac{m(A)}{H(A_l|A)^{n(A_l)}H(A_r|A)^{n(A_r)}}$ and $\eta_s(A)=\int r(A) \cdot \pi_{s}({\rm d}F(A_l|A))$. If $n(A)\leq 1$,  $r(A)=1$ by definition.

The recursion then terminates when $n(A)\leq 1$ with $\phi_s(A)=1$ in this case. Once the message passing has been completed either in terms of $\Phi_s(A)$ or $\phi_s(A)$, one can easily compute the corresponding posterior. Specifically, the posterior transition probability is then $\rho_{s,s'|\bx}(A):={\rm P}(S(A)=s'\,|\,S(A_p)=s,\bx) = \rho_{s,s'}(A)\eta_{s'}(A)\phi_{s'}(A_l)\phi_{s'}(A_r)/\phi_s(A).$
The conditional posterior for $F(A_l|A)$ given the latent state $S(A)=s$ is again simply given by the conjugate update $F(A_l|A)\,|\,S(A)=s,\bx \ind {\rm Beta}(\alpha_{s,l}(A)+n(A_l),\alpha_{s,r}(A)+n(A_r)).$

Under the partial likelihood framework and the data-dependent partition tree $\T$, the message passing algorithm in terms of either $\Phi_s(A)$ and $\phi_s(A)$ still applies, with the only difference being replacing the full likelihood on node $A$, $m(A)$, with the partial likelihood $m_{{\rm P}}(A)$, and the likelihood ratio $r(A)$ with $r_{{\rm P}}(A)$. The same posterior can be derived exactly in the same way with the same substitution.

There is a corresponding recursion for computing the posterior predictive density. As before, for any new point $\bx'\in \om$, let $\bx'\in A_k \subset A_{k-1} \subset \cdots \subset A_{0}=\om$ be the branch in the tree $\T$ which depends on the training data in which $\bx'$ belongs. Let us define $\xi_{s}(A):=\frac{{\rm E}(f(\bx'|A)|\bx,S(A_p)=s)}{h(\bx'|A)}$, which gives the posterior predictive density ratio conditional on $A$ given that $S(A_p)=s$. Then $\xi_{s}(A_k)=1$ for all $s$, and for $i<k$, $\xi_{s}(A_i)=\sum_{s'}\rho_{s,s'|\bx}(A)\cdot \varphi_{s'}(A_{i+1})\cdot \xi_{s'}(A_{i+1})$, where for all $s$ and all $A\in \T$,
$\varphi_{s}(A):=\frac{{\rm E}(F(A|A_{p})|\bx,S(A_p)=s)}{H(A|A_p)}$. Thus for $i<k$,
\begin{align*}
\varphi_{s}(A_{i+1}) &= \begin{cases}
\frac{\alpha_{s,l}(A_{i})+n(A_{i+1})}{\alpha_{s,l}(A_{i})+\alpha_{s,r}(A_{i)}+n(A_{i})-1} & \text{if $A_{i+1}$ is the left child of $A_i$}\\
\frac{\alpha_{s,r}(A_{i})+n(A_{i+1})}{\alpha_{s,l}(A_{i})+\alpha_{s,r}(A_{i})+n(A_{i})-1} & \text{if $A_{i+1}$ is the right child of $A_i$}.
 \end{cases}
\end{align*}

\noindent This provides a recursive recipe for computing the posterior predictive density ratio. Then the posterior predictive density is given by ${\rm E}(f(\bx')|\bx) = \xi_{0}(A_0) h(\bx') = \xi_{0}(\om) h(\bx').$ Note that because $A_0=\om$ has no parent, conditioning on the event that $S(A_p)=s$ for $A=\om$ is equivalent to not conditioning on anything.

While in computing the posterior one can proceed equivalently through either the marginal likelihood mapping $\Phi_s(A)$ or the likelihood-ratio mapping $\phi_s(A)$, for some models that incorporate a so-called stopping state under which $F(\cdot|A)=H(\cdot|A)$ w.p.1, one must remember that the partial likelihood under the stopping state should be the {\rm partial} likelihood under $H$, that is, excluding all of the partition locations within $A$. This is ensured automatically in the likelihood-ratio formulation. Similarly as before, one can precompute the beta functions involved in $\eta_s(A)=\int r_{\rm P}(A) \cdot \pi_{s}({\rm d}F(A_l|F))$ for all $s\in\S$ on a predefined grid of $(H(A_l|A),H(A_r|A))$ values as discussed before to speed up the computation. We provide numerical examples for density estimation using these models in Section~\ref{sec:num_ex_univariate}.

One can further incorporate adaptive partitioning to accommodate multivariate spaces using the partial likelihood following the OPT prior, which essentially imposes a Bayesian CART prior on the tree splitting. Details are provided in Supplementary Materials \hyperref[sec:supp_multivariate]{B}.

\section{Numerical experiments}
\label{sec:num_ex}

Next, we compare Bayesian inference using partial vs full likelihood under the OPT prior, assessing the performance in density estimation when using the OPT posterior mean as an estimator. The OPT is a PT-type model enhanced with a latent state variable $S(A)$ and endowed with a Bayesian CART prior on the tree structure in multivariate sample spaces. The latent state $S(A)$ takes values $0$ and $1$ and is modeled as Bernoulli$(\rho_{0,1}(A))$. The $S(A) = 1$ state describes a ``stopping'' state in which the prior within $A$ is degenerately concentrated on the base measure $H(\cdot|A)$, i.e. $F(\cdot|A)=H(\cdot|A)$.  

To make a fair comparison between the partial- and full- likelihood trees, the posterior predictive of the OPT is computed under both full and partial likelihoods with the same tree depth and hyperparameters. For simplicity and following the suggestion in \cite{wongandma2010},  we set the prior stopping probability equal to $\rho_{0, 1}(A) \equiv 0.5$ for all $A$, concentration parameters $c_{0}(A) = 2$ in the non-stopping case. 
For details on the computation of the posterior predictive in the univariate and bivariate examples, see  Supplementary Materials~\hyperref[sec:supp_exp]{C}.  
For each density estimation scenario, we assess the performance of the two models in terms of the average log $L_2$ risk (and standard error) across $200$ Monte Carlo experiments, for a given sample size and maximum tree depth. In a single Monte Carlo experiment, we compute the posterior predictive density of OPT for the full and partial-likelihood trees and the $L_2$ distance between the posterior density estimate and the true density. The $L_2$ distance strikes a balance between $L_1$ which is a global metric that focuses on large-scale features of the density and $L_{\infty}$ which is a local metric that focuses on local structures. Analogous results for $L_1$ and $L_{\infty}$ are reported in the Supplementary Materials \hyperref[sec:supp_exp]{C}.

\subsection{Univariate examples} 
\label{sec:num_ex_univariate}

We consider three cases of density estimation on the $[0, 1]$ interval. 
In the full likelihood tree, the left and right children of an interval are obtained by splitting over the midpoint of each interval. In the partial likelihood tree, the left and right children are obtained by splitting over the sample median of the observations in the interval (i.e., the interval observation of order $k = \lceil n\cdot p \rceil$, where $n$ is the number of points in the interval and $p = 0.5$). Moreover, as discussed in Section~\ref{sec:partial_like}, the data points coinciding with splitting locations are removed from the sample once splitting occurs (in the partial-likelihood tree). 

\begin{figure}[htb]
        \includegraphics[width=1\textwidth]{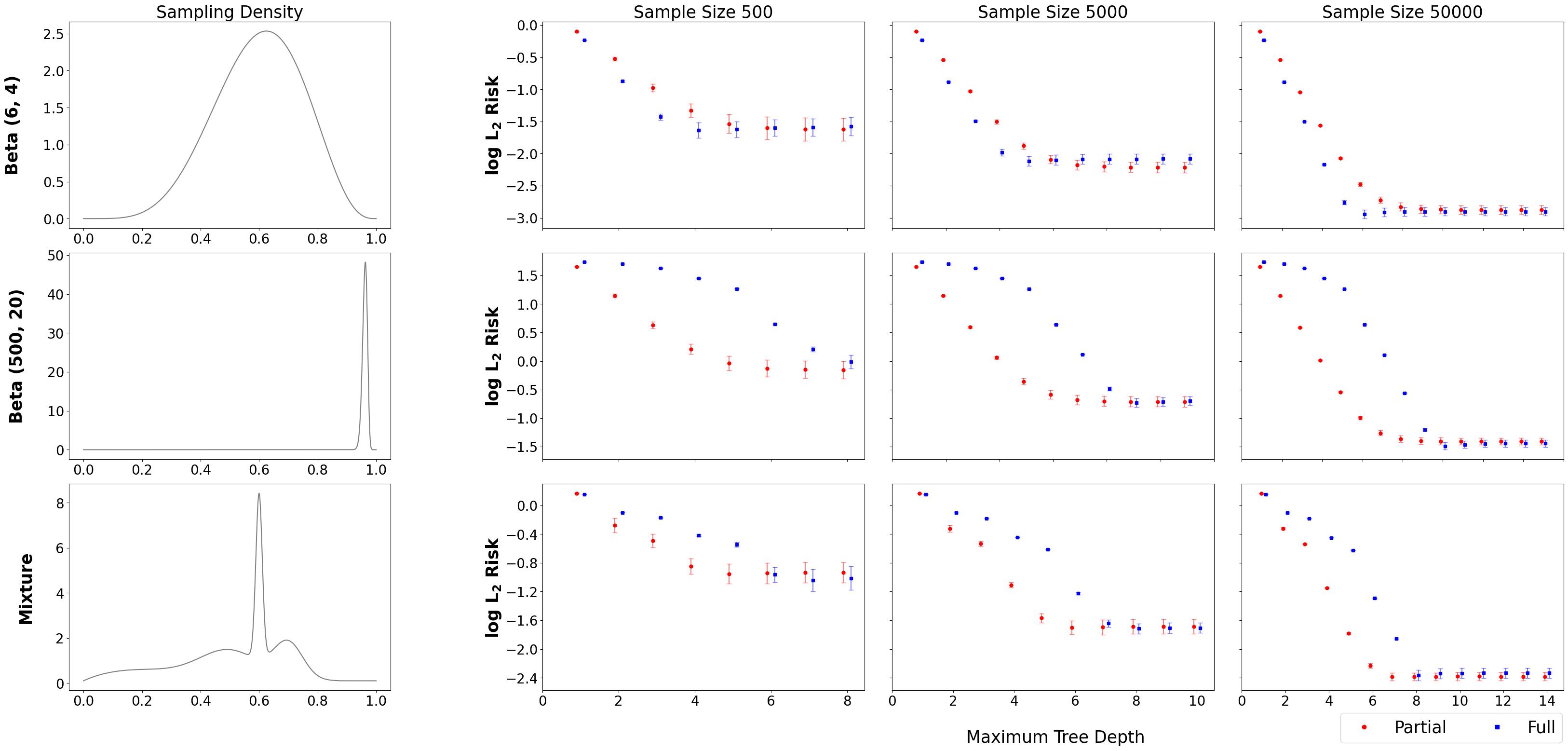}
        \caption{Boxplots for $L_2$ loss on univariate density estimation with OPT.}
        \label{fig:1D_L2}
\end{figure}

For each scenario, \ref{fig:1D_L2} reports the partial and full model average log $L_2$ risk (and standard error) for sample sizes varying in the range $[500, 5000, 50000]$ and maximum tree depth ranging from $1$ to at most $15$ (for similar results on $L_1$ and $L_{\infty}$, see \ref{fig:1D_L1_LI}).

At first glance, the partial- and full-likelihood trees show similar trends in the (log) $L_2$ risk, across all three density scenarios and sample sizes: the risk steeply decreases along the first few tree depths, to then stabilize in the medium to high range of resolutions. 
Moreover, the performance of both trees improves for larger sample sizes. However, we can notice different patterns when examining the three sampling densities in the low-to-medium range of depths.  In the Beta$(6, \:4)$ case, the full-likelihood tree shows lower risk than the partial-likelihood tree in this depth range. This suggests that for such a smooth and rather symmetric sampling density, the even splitting employed in the full-likelihood tree is roughly the ``right'' partition to adopt. Such is not the case when the  density is less even over the sample space. For example, in the Beta$(500, \:20)$ case, the $L_2$ risk of the partial-likelihood tree decreases much more rapidly and tends to plateau earlier than in the full-likelihood tree. 

The partial-likelihood tree also dominates the full-likelihood one for such depth range for a mixture scenario with heterogeneous smoothness over the sample space:
\begin{align*}
0.1*{\rm Uniform}(0, \:1) \:+ 0.2*{\rm Beta}(2, \:5) \:+ 0.2*{\rm Beta}(1200, \:800) \\
\:+ 0.3*{\rm N}_{[0.1, \:0.9]}(0.5, 0.1^2) \:+ 0.2*{\rm N}_{[0.3, \:0.87]}(0.7, 0.05^2)
\end{align*}
where ${\rm N}_{[a,b]}$ represents a normal distribution truncated into the interval $[a,b]$. Eventually, of course, for deep enough trees, the full-likelihood tree catches up, but it requires a much deeper tree and hence incurs additional computational costs. This is consistent with our intuition—the dominating component in the risk at low resolutions is the bias or inflexibility of the model, but as the model grows richer, eventually the bias is no longer important and the adaptive protects the variance (or overfitting) from dominating.

\subsection{Bivariate examples}
\label{sec:num_ex_bivariate}
We present results for a set of bivariate density estimation problems on the unit rectangle $[0,1] \times [0, 1]$. In these examples, the dyadic partitioning process can split a set on either dimension, randomly selected with a prior probability of $0.5$. Once the dimension is selected, the full and partial likelihood trees differ in splitting locations. For the full likelihood tree, a set is partitioned on its marginal midpoint in the chosen dimension. For the partial likelihood tree, a set is partitioned over the marginal sample median of the set points in the chosen dimension (the partition location is the data point having dimension coordinate of order $k = \lceil n\cdot p \rceil$, where $n$ is the number of points in the set and $p = 0.5$). In the experiments below, when partitioning a node, we condition on the sample medians of both dimensions following the second strategy of choosing the partition points discussed in Section~\ref{sec:multivar}. 

\begin{figure}[htbp]
        \hspace{-1cm}
        \includegraphics[width=1.1\textwidth]{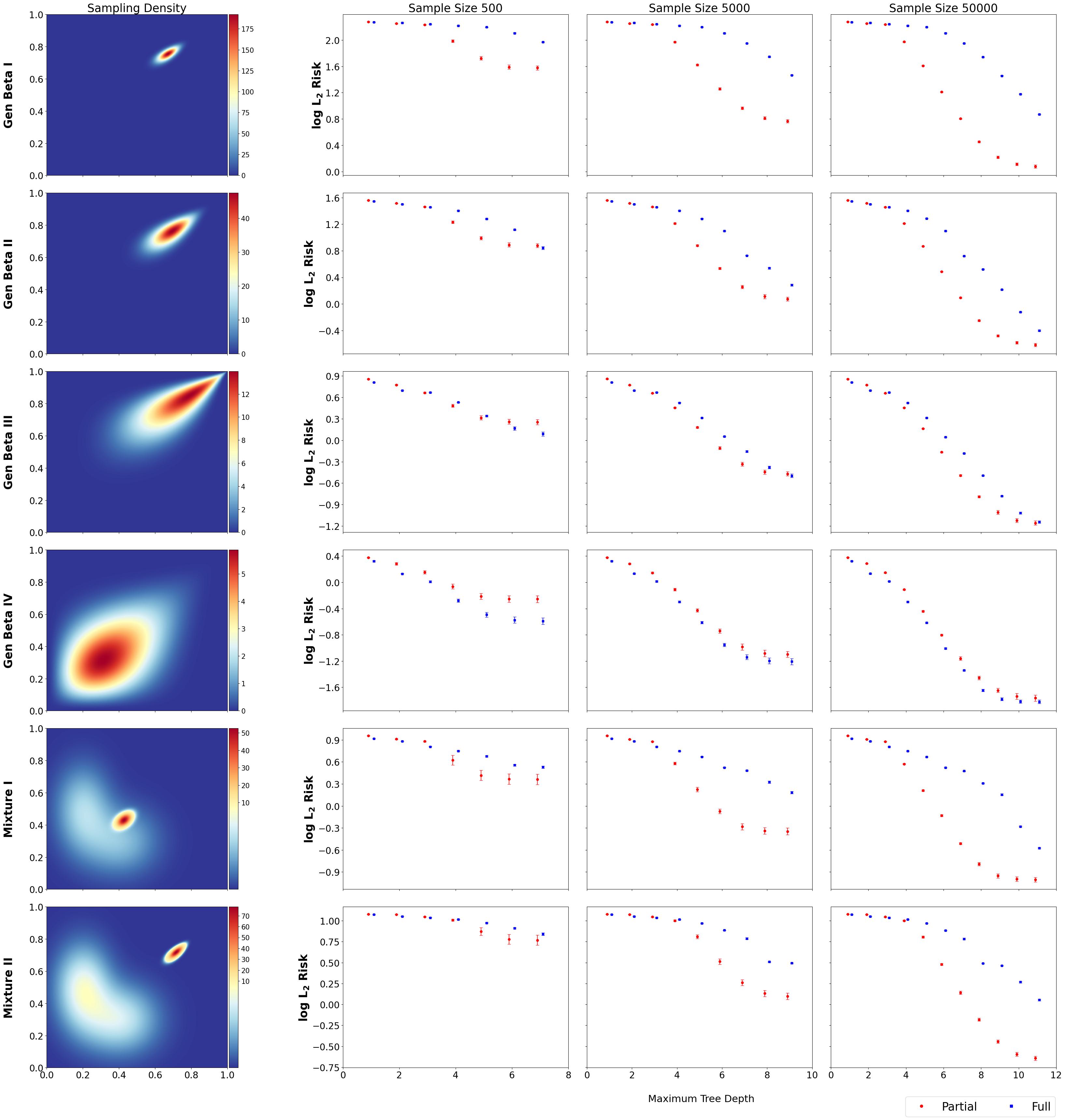}
        \caption{$Log \: L_2$ simulation results for bivariate Optional Polya Trees. }
        \label{fig:2D_L2}
\end{figure}

We consider six density estimation scenarios in the unit rectangle, represented in the ``Sampling Density'' column in \ref{fig:2D_L2}: 
\vspace{0.5em}

\noindent \textit{Generalized Beta I:} with parameters $(\alpha_0, \beta_0, \alpha_1, \beta_1, \alpha_2, \beta_2) = (50, 1, 100, 1, 150, 1);$ \\
\textit{Generalized Beta II:} with parameters $(\alpha_0, \beta_0, \alpha_1, \beta_1, \alpha_2, \beta_2) = (12, 1, 25, 1, 35, 1)$;\\
\textit{Generalized Beta III:} with parameters $(\alpha_0, \beta_0, \alpha_1, \beta_1, \alpha_2, \beta_2) = (3, 1, 6, 1, 9, 1)$;\\
\textit{Generalized Beta IV:} with parameters $(\alpha_0, \beta_0, \alpha_1, \beta_1, \alpha_2, \beta_2) = (5, 10, 3, 10, 3, 10)$; \\
\noindent \textit{Mixture I}: $0.4*{\rm N}_{[0, 1] \times [0,1]} (\mu_{a}, \Sigma_a)  +  0.4*{\rm N}_{[0, 1] \times [0,1]}  (\mu_{b}, \Sigma_b) + {\rm GBeta} (200, 1, 150, 1, 150, 1);$ \\
\noindent \textit{Mixture II}: $0.4*{\rm N}_{[0, 1] \times [0,1]} (\mu_{a}, \Sigma_a)  +  0.4*{\rm N}_{[0, 1] \times [0,1]}  (\mu_{b}, \Sigma_b) + {\rm GBeta} (100, 1, 250, 1, 250, 1)$,
\vspace{0.5em}

\noindent where ${\rm N}_{[0, 1] \times [0,1]}$ is a Normal distribution truncated over the unit interval, $\mu_a = [0.2, 0.5]^T$,  $\mu_b = [0.4, 0.3]^T$, $\Sigma_a = diag([0.01, 0.03])$, $\Sigma_b = diag([0.02, 0.02])$, and $GBeta$ a Generalized Beta density. (More details on this density can be found in the Supplementary Materials \hyperref[sec:supp_exp]{C}.)

For each density estimation scenario, in \ref{fig:2D_L2}, we report the log $L_2$ risk across 200 Monte Carlo experiments, for sample sizes 500, 5000, and 50000. We consider maximum tree depth ranging between $1$ and $11$, depending on the sample size. We also report results for $L_1$, $L_{\infty}$ in \ref{fig:2D_L1_LI} in the Supplementary Materials.

\begin{figure}[htbp]
    
        \hspace{-1cm}
        \centering
        \includegraphics[width=0.7\textwidth]{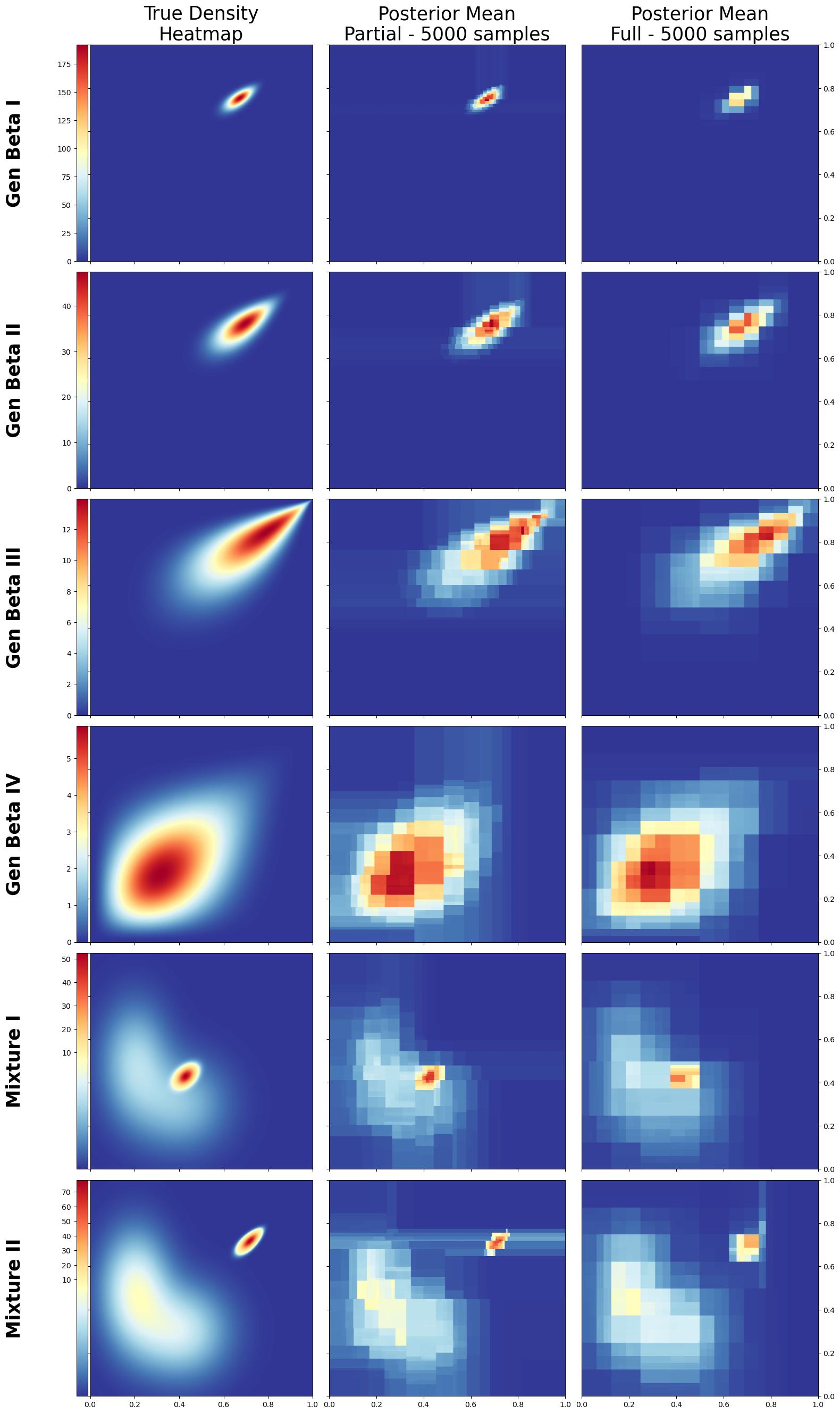}
        \caption{The posterior predictive density of a bivariate OPT, at tree depth $9$ for sample size $5000$ under six simulation scenarios. Left: True density; Middle: PPD with partial likelihood; Right: PPD with full likelihood. Each row corresponds to a simulation scenario.}
        \label{fig:2D_post_mean}
\end{figure}

The simulation results for generalized beta densities confirm the findings in the univariate examples. For both tree models, the risk tends to decrease at higher depths and larger sample sizes, especially for $L_1$ and $L_2$ risks. Generally, the partial-likelihood model is preferable to the full-likelihood model in the presence of higher density spikes. This trend is confirmed for the most extreme case, ``Generalized Beta~I'', where the partial model outperforms the full model for all sample sizes with deep enough trees. The more gentle spikes of ``Generalized Beta II'' and ``III'' still illustrate a relative advantage of the partial model maintained for intermediate to high resolution levels, and especially for $L_{2}, L_{\infty}$, metrics that better capture local discrepancies. However, the full-likelihood tree catches up with the partial-likelihood tree with smoother sampling densities, as illustrated by the results for  ``Generalized Beta IV''. The performance of the partial model under the different Generalized Betas suggests that the full-likelihood model performs reasonably for smooth and symmetric sampling densities. However, the partial-likelihood tree is preferable when the true density is locally concentrated and/or skewed.  
The partial-likelihood tree still outperforms the full-likelihood tree in both mixture examples, despite the presence of smooth low-resolution structures.

A notable observation is that the full-likelihood tree never catches up with the partial-likelihood tree up to the maximum depth of 10, at which point the risk of the partial-likelihood tree has already plateaued for all of the scenarios. The fact that this occurs even in just a two-dimensional sample space is striking. The gain from data-adaptive design of the sampling model becomes ever more substantial as the dimensionality grows in that the computational cost associated with a data-agnostic partition system can become prohibitive.

Finally, \ref{fig:2D_post_mean} displays some examples of posterior mean under the full and partial models for a sample size of $5000$ and tree depth of $9$,  which shows that the large bias or underfitting under the full-likelihood tree leads to severe pixelation in the estimated density. The partial likelihood-based estimate is much more effective in capturing the heterogeneous smoothness of the underlying density.

\subsection{An example from flow cytometry}

We demonstrate the application of the partial likelihood approach to density estimation on a data set from a flow cytometry experiment. In flow cytometry, multiple fluorescence and physical markers are measured for individual cells in a blood sample. 
The values of the markers are used to classify cells into subtypes through a process called ``gating". This process involves identifying cell groups by drawing boundaries on histogram plots of pairs of markers of interest. Subgroup identification is typically performed by visual inspection of 2D projections of the cell observations, which can be a time-consuming process. Some algorithmic approaches  have been developed to replace this manual gating process in an automatic, high-throughput fashion \citep{cheung2021current}. A common strategy in these automatic gating methods is through mode-hunting strategies based on an estimated density function for 2D projections of the data on pairs of markers \citep{waltherandzhao2023}.

\begin{figure}[htbp]
        \hspace{-1cm}
        \centering
        \includegraphics[width=0.85\textwidth]{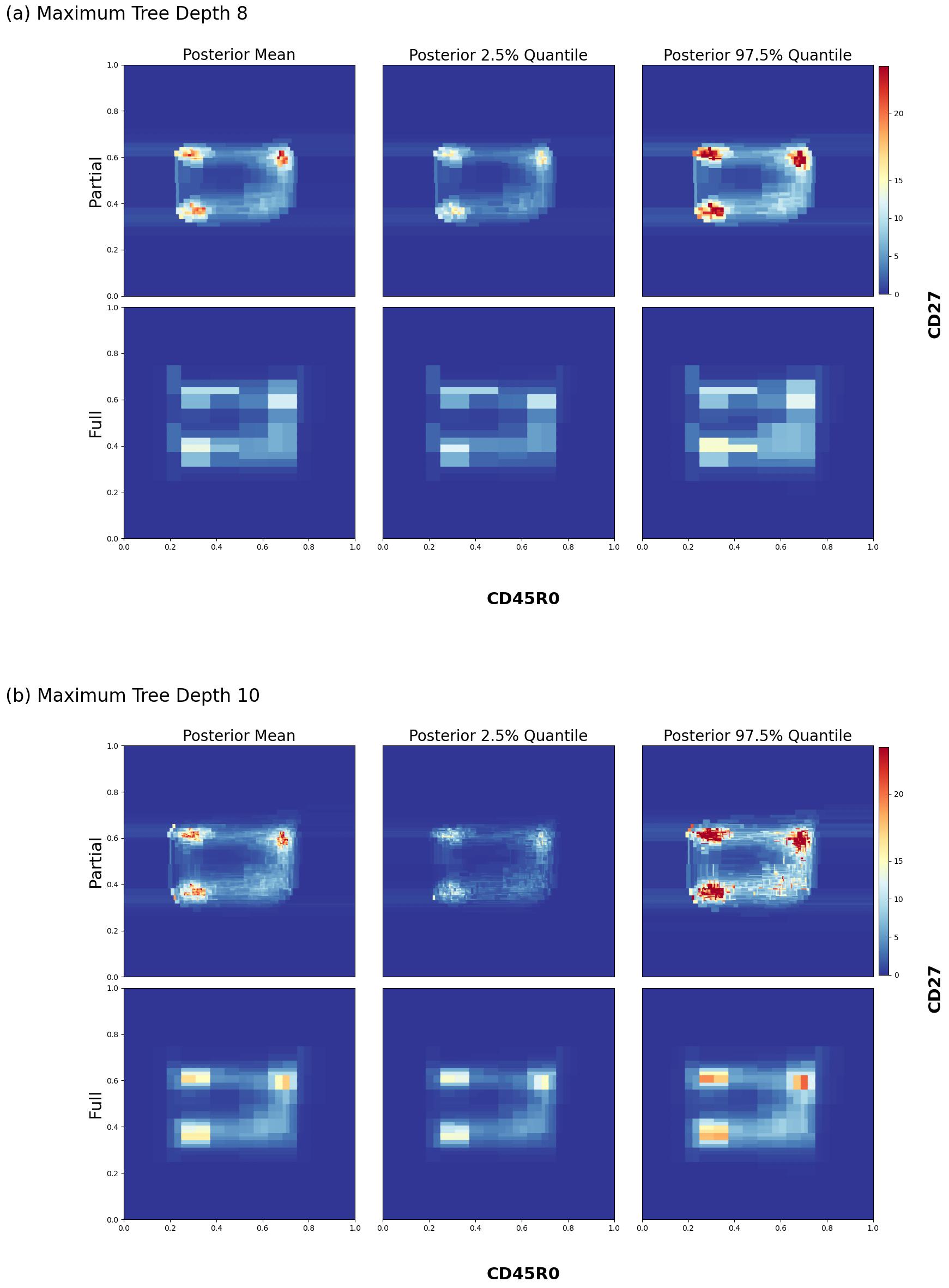}
        \caption{Estimated density on two markers---CD45RO and CD$27$---for the flow cytometry data on a sample of $10000$ cells and 95\% credible bands: (a) PPD from trees with depth 8; (b) PPD from trees with depth 10. In each panel, first row: results for partial-likelihood; second row: results for full-likelihood. Left: PPD; middle: $2.5\%$ pointwise posterior quantile; right: $97.5\%$ pointwise posterior quantile.}
        \label{fig:2D_cd45_cd27}
\end{figure}

We apply tree-based density estimation on a sample of $10000$ blood cells from a flow cytometry dataset collected at Duke University \citep{staats2014toward}. The cells are measured on $14$ markers. The density for each marker pair is estimated with partial- and full-likelihood OPT. \ref{fig:2D_cd45_cd27} reports the posterior mean densities with tree depths 8 and 10 for markers CD45RO and CD27, along with a pointwise 95\% credible band given by the $2.5\%$ and $97.5\%$ posterior quantiles, all computed using $10000$ Monte Carlo posterior samples. 
CD45RO and CD27 markers can be used to identify subsets of T cells for their memory and effector characteristics. In particular, they can help distinguish four major subsets  \citep{maecker2012standardizing}:  CD27$^+$  CD45RO$^+$ T cells correspond to central memory T cells,  CD27$^-$  CD45RO$^+$ correspond to effector memory T cells while CD27$^+$  CD45RO$^-$ represent naïve T cells and CD27$^-$  CD45RO$^-$ are terminally differentiated effector-memory T cells. 

The partial-likelihood posterior mean and credible band in \ref{fig:2D_cd45_cd27} suggest the presence of the four cell clusters at just depth $8$. 
In contrast, the full-likelihood tree at depth $8$ fails to clearly distinguish four cell clusters. Indeed, the full-likelihood tree needs a deeper partition (depth $10$) to further differentiate the groups, and still does not offer the same level of resolution as the partial-likelihood at only depth~8.

\section{Discussion}
We have introduced a framework for incorporating partition locations that sit on observed data values into tree-based density modeling and demonstrated its use in the context of Bayesian inference on such models. Using a small amount of information from the data to help design the sampling model can drastically improve the parsimony of the model in effectively representing the underlying distribution. The partial likelihood provides a principled way to achieve consistent inference without inventing new tree-based priors.

Beyond dyadic partitions that divide always at a given quantile, one can further incorporate selection variables that choose also the partition location. The partition location can then be on each of the observed values of the training data along a dimension. The notation becomes more complicated because then the $k$th order statistic is no longer a simple function of $A$ but also a function of $(j,l)$, where $j$ is the dimension to divide and $l$ is the location to divide. In that case, we can now condition on the $k_{j,l}(A)$th order statistic. An important implication of this is that the partial likelihood approach can be applied to allow possible partition on each of the observed values along each dimension in each node $A$, giving a probabilistic version of the non-Bayesian CART partition strategy. Exact recursive calculation of the corresponding posterior will be computationally intractable for multivariate problems with even just a few hundred data points. Simplification could be made to consider just a subset of observed values as candidate partition locations.

\cite{lu2013} showed that one could effectively sample from the posterior tree model when there are a large number of predetermined partition locations in each dimension of a multivariate sample space using a sequential Monte Carlo (SMC) strategy. \cite{awayaandma2023} further showed that this SMC still works even when latent states are incorporated. Future work will examine how the SMC sampling strategy works in the partial likelihood framework. 
This will allow us to restrict the partition locations on one or multiple of the actual observed data values in each direction, thereby substantially reducing the number of particles needed and improving computational efficiency.

\subsection*{Acknowledgment}

This research is partly supported by NSF grants DMS-1749789 and DMS-2152999, as well
as NIGMS grant R01-GM135440. The authors thank Cliburn Chan for helpful comments.

\bibliographystyle{Chicago}

\bibliography{references}

\begin{thebibliography}{}

\bibitem[\protect\citeauthoryear{Awaya and Ma}{Awaya and
  Ma}{2024}]{awayaandma2023}
Awaya, N. and L.~Ma (2024).
\newblock Hidden markov pólya trees for high-dimensional distributions.
\newblock {\em Journal of the American Statistical Association\/}~{\em
  119\/}(545), 189--201.

\bibitem[\protect\citeauthoryear{Berger and Wolpert}{Berger and
  Wolpert}{1988}]{berger1988likelihood}
Berger, J. and R.~Wolpert (1988).
\newblock {\em The Likelihood Principle}.
\newblock Institute of Mathematical Statistics. Lecture notes : monographs
  series. Institute of Mathematical Statistics.

\bibitem[\protect\citeauthoryear{Berger and Guglielmi}{Berger and
  Guglielmi}{2001}]{bergerandguglielmi2001}
Berger, J.~O. and A.~Guglielmi (2001).
\newblock Bayesian and conditional frequentist testing of a parametric model
  versus nonparametric alternatives.
\newblock {\em Journal of the American Statistical Association\/}~{\em
  96\/}(453), 174--184.

\bibitem[\protect\citeauthoryear{Cheung, Campbell, Whitby, Thomas, Braybrook,
  and Petzing}{Cheung et~al.}{2021}]{cheung2021current}
Cheung, M., J.~J. Campbell, L.~Whitby, R.~J. Thomas, J.~Braybrook, and
  J.~Petzing (2021).
\newblock Current trends in flow cytometry automated data analysis software.
\newblock {\em Cytometry Part A\/}~{\em 99\/}(10), 1007--1021.

\bibitem[\protect\citeauthoryear{Chipman, George, and McCulloch}{Chipman
  et~al.}{1998}]{chipman1998}
Chipman, H.~A., E.~I. George, and R.~E. McCulloch (1998).
\newblock Bayesian cart model search.
\newblock {\em Journal of the American Statistical Association\/}~{\em
  93\/}(443), 935--948.

\bibitem[\protect\citeauthoryear{Christensen and Ma}{Christensen and
  Ma}{2019}]{christensenandma2020}
Christensen, J. and L.~Ma (2019, 11).
\newblock A bayesian hierarchical model for related densities by using pólya
  trees.
\newblock {\em Journal of the Royal Statistical Society Series B: Statistical
  Methodology\/}~{\em 82\/}(1), 127--153.

\bibitem[\protect\citeauthoryear{Cox}{Cox}{1972}]{cox1972}
Cox, D.~R. (1972).
\newblock Regression models and life-tables.
\newblock {\em Journal of the Royal Statistical Society: Series B
  (Methodological)\/}~{\em 34\/}(2), 187--202.

\bibitem[\protect\citeauthoryear{Cox}{Cox}{1975}]{cox1975}
Cox, D.~R. (1975).
\newblock Partial likelihood.
\newblock {\em Biometrika\/}~{\em 62\/}(2), 269--276.

\bibitem[\protect\citeauthoryear{Ferguson}{Ferguson}{1973}]{ferguson1973}
Ferguson, T.~S. (1973).
\newblock A bayesian analysis of some nonparametric problems.
\newblock {\em The annals of statistics\/}, 209--230.

\bibitem[\protect\citeauthoryear{Freedman}{Freedman}{1963}]{freedman1963}
Freedman, D.~A. (1963).
\newblock On the asymptotic behavior of bayes' estimates in the discrete case.
\newblock {\em The Annals of Mathematical Statistics\/}~{\em 34\/}(4),
  1386--1403.

\bibitem[\protect\citeauthoryear{Hjort and Walker}{Hjort and
  Walker}{2009}]{hjortandwalker2009}
Hjort, N.~L. and S.~G. Walker (2009).
\newblock Quantile pyramids for bayesian nonparametrics.
\newblock {\em The Annals of Statistics\/}~{\em 37\/}(1), 105--131.

\bibitem[\protect\citeauthoryear{Holmes, Caron, Griffin, and Stephens}{Holmes
  et~al.}{2015}]{holmes2015}
Holmes, C.~C., F.~Caron, J.~E. Griffin, and D.~A. Stephens (2015).
\newblock Two-sample bayesian nonparametric hypothesis testing.
\newblock {\em Bayesian Analysis\/}~{\em 10\/}(2), 297--320.

\bibitem[\protect\citeauthoryear{Lavine}{Lavine}{1992}]{lavine1992}
Lavine, M. (1992).
\newblock Some aspects of polya tree distributions for statistical modelling.
\newblock {\em The Annals of Statistics\/}~{\em 20\/}(3), 1222--1235.

\bibitem[\protect\citeauthoryear{Lavine}{Lavine}{1995}]{lavine1995}
Lavine, M. (1995).
\newblock On an approximate likelihood for quantiles.
\newblock {\em Biometrika\/}~{\em 82\/}(1), 220--222.

\bibitem[\protect\citeauthoryear{Lu, Jiang, and Wong}{Lu et~al.}{2013}]{lu2013}
Lu, L., H.~Jiang, and W.~H. Wong (2013).
\newblock Multivariate density estimation by bayesian sequential partitioning.
\newblock {\em Journal of the American Statistical Association\/}~{\em
  108\/}(504), 1402--1410.

\bibitem[\protect\citeauthoryear{Ma}{Ma}{2017}]{ma2017}
Ma, L. (2017).
\newblock {Adaptive Shrinkage in Pólya Tree Type Models}.
\newblock {\em Bayesian Analysis\/}~{\em 12\/}(3), 779 -- 805.

\bibitem[\protect\citeauthoryear{Ma and Wong}{Ma and
  Wong}{2011}]{maandwong2011}
Ma, L. and W.~H. Wong (2011).
\newblock Coupling optional pólya trees and the two sample problem.
\newblock {\em Journal of the American Statistical Association\/}~{\em
  106\/}(496), 1553--1565.

\bibitem[\protect\citeauthoryear{Maecker, McCoy, and Nussenblatt}{Maecker
  et~al.}{2012}]{maecker2012standardizing}
Maecker, H.~T., J.~P. McCoy, and R.~Nussenblatt (2012).
\newblock Standardizing immunophenotyping for the human immunology project.
\newblock {\em Nature Reviews Immunology\/}~{\em 12\/}(3), 191--200.

\bibitem[\protect\citeauthoryear{Soriano and Ma}{Soriano and
  Ma}{2017}]{sorianoandma2017}
Soriano, J. and L.~Ma (2017).
\newblock Probabilistic multi-resolution scanning for two-sample differences.
\newblock {\em Journal of the Royal Statistical Society Series B: Statistical
  Methodology\/}~{\em 79\/}(2), 547--572.

\bibitem[\protect\citeauthoryear{Staats, Enzor, Sanchez, Rountree, Chan,
  Jaimes, Chan, Gaur, Denny, and Weinhold}{Staats
  et~al.}{2014}]{staats2014toward}
Staats, J.~S., J.~H. Enzor, A.~M. Sanchez, W.~Rountree, C.~Chan, M.~Jaimes,
  R.~C.-F. Chan, A.~Gaur, T.~N. Denny, and K.~J. Weinhold (2014).
\newblock Toward development of a comprehensive external quality assurance
  program for polyfunctional intracellular cytokine staining assays.
\newblock {\em Journal of immunological methods\/}~{\em 409}, 44--53.

\bibitem[\protect\citeauthoryear{Walther and Zhao}{Walther and
  Zhao}{2023}]{waltherandzhao2023}
Walther, G. and Q.~Zhao (2023).
\newblock Beta-trees: Multivariate histograms with confidence statements.

\bibitem[\protect\citeauthoryear{Wong and Ma}{Wong and
  Ma}{2010}]{wongandma2010}
Wong, W.~H. and L.~Ma (2010).
\newblock {Optional Pólya tree and Bayesian inference}.
\newblock {\em The Annals of Statistics\/}~{\em 38\/}(3), 1433 -- 1459.

\end{thebibliography}

\newpage 

\clearpage
\renewcommand{\thepage}{S\arabic{page}} 
\setcounter{page}{1}

\renewcommand{\thefigure}{S\arabic{figure}} 
\setcounter{figure}{0}

\renewcommand{\thesection}{\Alph{section}} 
\setcounter{section}{0}

\section{Bayesian modeling based on the partial likelihood for multivariate distributions}
\label{sec:supp_multivariate}

Next we demonstrate how to incorporate adaptive partitioning in multivariate spaces using the partial likelihood following the OPT prior, which essentially imposes a Bayesian CART prior on the tree splitting. 
Specifically, for each potential tree node $A$, we adopt a multinomial prior  on $J(A)$ over the available dimensions to divide. That is, $J(A)\in \{1,2,\ldots,d\}$ is a multinomial Bernoulli variable with probability $\blam(A)=(\lambda_1(A),\lambda_2(A),\ldots,\lambda_d(A))$ where $0\leq \lambda_j(A)\leq 1$ for all $j$ and $\sum_{j}\lambda_j(A)=1$. Given that $J(A)=j$, then we divide $A$ in the $j$th dimension into the two children $A_{j,l}$ and $A_{j,r}$ and the splitting variable $J(A_{j,l})$ and $J(A_{j,r})$ are further generated in the same way. The partial likelihood of $(f,\bJ)$ then follows the expression in Eq.~\eqref{eq:likelihood_multi_partial}. 

We can again express the partial likelihood in a scale-free fashion in terms of the likelihood-ratio with respect to a base measure $H$. Let 
\[
r_j(A)=\frac{m_{{\rm P},j}(A)}{H(A_{j,l}|A)^{n(A_{j,l})} H(A_{j,r}|A)^{n(A_{j,r})}}.
\]
Then for $J(A)=j$,
\[
\frac{L_{\rm P}(f,\bJ;\bx,A)}{L_{\rm P}(h,\bJ;\bx,A)}=
r_j(A) \frac{L_{\rm P}(f,\bJ;\bx,A_{j,l})}{L_{\rm P}(h,\bJ;\bx,A_{j,l})}\frac{L_{\rm P}(f,\bJ;\bx,A_{j,r})}{L_{\rm P}(h,\bJ;\bx,A_{j,r})}.
\]
where for each $j$ $L_{\rm P}(h;\bx,A)=H(A_{j,l}|A)^{n(A_{j,l})} H(A_{j,r}|A)^{n(A_{j,r})}\cdot L_{\rm P}(h;\bx,A_{j,l})\cdot L_{\rm P}(h;\bx,A_{j,r}).$

Moreover, one can compute the exact posterior distribution of $\bJ$ by integrating out the conditional prior on $f$, such as the PT, as follows. Let 
$\Phi(A)=\int L_{\rm P}(f;\bx,\bJ,A)\pi(df,d\bJ)$.
Then $\Phi(A)$ can be computed recursively 
\begin{align*}
\Phi(A)&=\int\int L_{\rm P}(f;\bx,\bJ,A)\pi(df|\bJ)\pi(d\bJ)=\sum_{j} \lambda_j(A)M_j(A) \Phi(A_{j,l})\Phi(A_{j,r})
\end{align*}
where $M_j(A)=\int m_{j}(A) \pi({\rm d}F(A_{j,l}|A)).$
Again, the integral term in the middle is simply a beta function when beta priors are adopted on $F(A_{j,l}|A)$. Correspondingly, the posterior probability for $J(A)=j$ given the training data $\bx$ is ${\rm P}(J(A)=j|\bx)=\lambda_j(A)M_j(A)\cdot \Phi(A_{j,l})\Phi(A_{j,r})/\Phi(A).$

The conditional posterior for $f$ given $\bJ$ is exactly the same as in the case with a predetermined partition tree. One can similarly write the above recursion in a normalized fashion in terms of the likelihood ratio with respect to the base density $\phi(A)=\Phi(A)/\prod_{x\in A}h(x|A)$. In which case $\phi(A)=\sum_{j} \lambda_j(A)\eta_j(A) \phi(A_{j,l})\phi(A_{j,r})$ where \newline 
$\eta_j(A)=\int r_j(A)\pi(df,d\bJ)= M_j(A)/H(A_{j,l}|A)^{n(A_{j,l})} H(A_{j,r}|A)^{n(A_{j,r})}$. Accordingly, \newline ${\rm P}(J(A)=j|\bx)=\lambda_j(A)\eta_j(A)\cdot \phi(A_{j,l})\Phi(A_{j,r})/\phi(A).$

Furthermore, one can specify the conditional model on $f$ given the tree using latent state variables. When the underlying latent states are modeled using a Markov process, for example, one can still follow the recipe described previously and arrive at the following expression for the marginal likelihood on $A$, after integrating out both the state variables as well as the splitting variables
\begin{align}
\Phi_{s}(A)&=\sum_{j}\sum_{s'\in \S}\lambda_{j}(A)\rho_{s,s'}(A)M_{j,s'}(A)\Phi_{s'}(A_{j,l})\Phi_{s'}(A_{j,r}).
\end{align}
where for all $s$, $M_{j,s}(A)=\int m_{{\rm P},j}(A) \cdot \pi_{s}({\rm d}F(A_l|A))$. Intuitively, $\Phi_s(A)$ is the marginal likelihood for $A$ given that $S(A_p)=s$.

Even though we had specified the priors for $\bS$ and $\bJ$ independently, {\em a posteriori} they are dependent. So we must describe the joint posterior of $(\bJ,\bS)$, which is jointly a Markov process. Specifically, the  posterior transition probability for $(J(A),S(A))$ is
\begin{align*}
{\rm P}(J(A)=j',S(A)=s'|J(A_p)=j,S(A_p)=s,\bx)=&\lambda_{j'}(A)\rho_{s,s'}(A)M_{j',s'}\Phi_{s'}(A_{j',l})\Phi_{s'}(A_{j',r})/\Phi_s(A).
\end{align*}
Note that the dependence on $J(A_p)=j$ is implicit as it determines the node $A$.
Given the values of $(\bJ,\bS)$, the conditional posterior of $F$ boils down to the standard univariate PT. 

All of the above expressions can be written in terms of the likelihood ratio with respect to the base density by replacing $\Phi_s(A)$ with $\phi_{s}(A)=\Phi_s(A)/\prod_{x\in A}h(x|A)$ and $M_{j,s}(A)$ with $\eta_{j,s}(A)=M_{j,s}(A)/\left(H(A_{j,l}|A)^{n(A_{j,l})}H(A_{j,r}|A)^{n(A_{j,r})}\right)$. That is
\begin{align}
\phi_{s}(A)&=\sum_{j}\sum_{s'\in \S}\lambda_{j}(A)\rho_{s,s'}(A)\eta_{j,s'}(A)\phi_{s'}(A_{j,l})\phi_{s'}(A_{j,r})
\end{align}
and
\begin{align}
{\rm P}(J(A)=j',S(A)=s'|J(A_p)=j,S(A_p)=s,\bx)=&\lambda_{j'}(A)\rho_{s,s'}(A)\eta_{j',s'}\phi_{s'}(A_{j',l})\phi_{s'}(A_{j',r})/\phi_s(A).
\end{align}

This gives a complete characterization of the marginal posterior of $(\bJ,\bS)$. Given $(\bJ,\bS)$, the conditional posterior of $F$ is simply that of the corresponding PT model. We provide numerical examples for inference on multivariate sample spaces in Section~\ref{sec:num_ex_bivariate}.

\newpage

\section{Additional details and results of the numerical experiments}
\label{sec:supp_exp}

We report additional numerical results for the univariate and multivariate simulations previously discussed.

\subsection{Univariate OPT}
\label{sec:supp_univ}

\subsubsection{OPT PPD Computation}

The OPT posterior mean is computed exactly for univariate examples with the recursive algorithm for latent-state models in Section~\ref{sec:latent_state}. 

\subsubsection{Results for $L_1, L_{\infty}$}

\begin{figure}
        \hspace{-1cm}
        \includegraphics[width=1.1\textwidth]{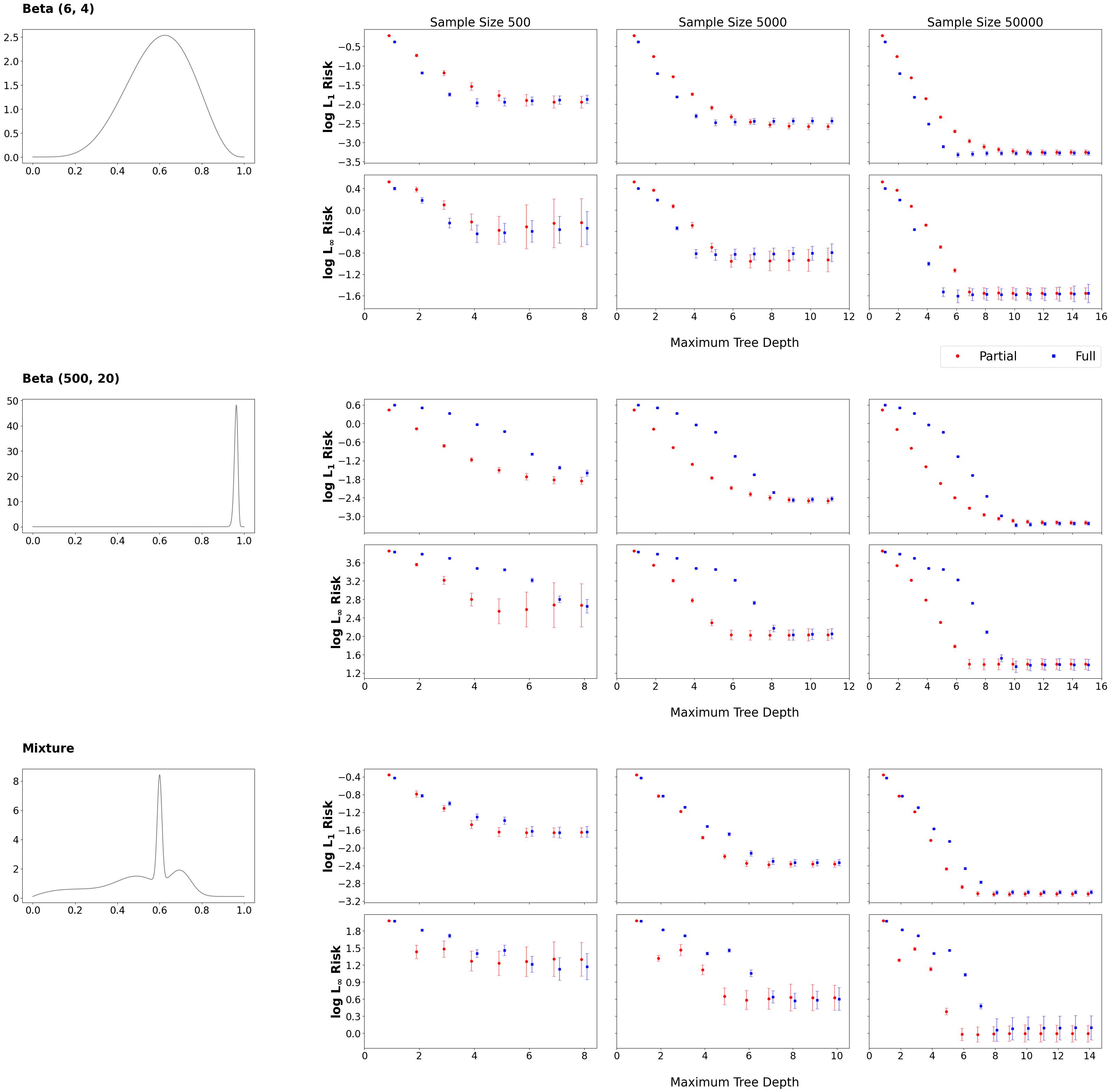}
        \caption{Simulation results for log $L_1, L_{\infty}$ risks in Univariate examples. }
        \label{fig:1D_L1_LI}
\end{figure}

\ref{fig:1D_L1_LI} reports the log $L_1, L_{\infty}$ risks obtained under the same experiments discussed in Section~\ref{sec:num_ex_univariate}. Both metrics generally suggest the same patterns already observed in the $L_2$ risk: a steep decrease in the first few tree depths and then plateau around some value, with the partial-likelihood tree having better performance than the full-likelihood tree in the low to medium depth for the $Beta(500, 20)$ and $Mixture$ cases. However, these patterns are not as regular as in $L_2$, especially when considering a sample size of $500$ and the $L_{\infty}$ metric, which, having a more local nature, is more sensitive to lower sample sizes than the other two metrics. \newline 

\subsection{Bivariate OPT}
\label{sec:supp_biv}

\subsubsection{Construction of the Generalized Beta Density. }

The generalized beta densities used in the bivariate density estimation problems were constructed as follows. Let $G_0 \sim {\rm Gamma}(\alpha_0, \beta_0)$, $G_1 \sim {\rm Gamma}(\alpha_1, \beta_1)$ and $G_2 \sim {\rm Gamma}(\alpha_2, \beta_2)$ be three independent gamma random variables, with $\alpha_1, \beta_i > 0$ for $i = 0, 1, 2$. 
The joint density of $X = G_1 / (G_1 + G_0)$ and $Y = G_2 / (G_2 + G_0)$, with $x, y \in (0, 1)$ has the form 
\begin{align*}
    f(x, y) = \frac{1}{B(\alpha_0, \alpha_1, \alpha_2)} \frac{\lambda_1^{\alpha_1} x^{\alpha_1 -1} (1-x)^{-(\alpha_1 +1)}\lambda_2^{\alpha_2} y^{\alpha_2 -1} (1-y)^{-(\alpha_2 +1)}}{[1 + \lambda_1 x/(1-x) + \lambda_2 y / (1-y)]^{\alpha_0 + \alpha_1 + \alpha_2}}
\end{align*}
and $\lambda_i = \beta_i/\beta_0$ for $i = 1, 2$ and $ B(\alpha_0, \alpha_1, \alpha_2) = \prod_{i = 0, 1, 2} \Gamma(\alpha_i)/\Gamma(\sum \alpha_i))$. 

\subsubsection{OPT PPD Computation}

In the bivariate OPT examples, we compute the posterior mean through Monte Carlo samples of the trees. This method is straightforward to implement in a setting where randomizing the splitting dimension for a set requires accounting for all possible children generated by that set.
For each MC run, we first compute the message passing recursion terms as in Section~\ref{sec:latent_state} for all possible sets generated through all possible trees, to compute posterior stopping probabilities and splitting probabilities along each dimension. Next, we sample a tree using a top-down algorithm: for an available set, we sample a latent state (determining stopping and non-stopping) with the posterior stopping probability and next we sample a splitting dimension using the posterior splitting probability conditional on the sampled state and compute its children. Given such a tree, we evaluate the posterior mean on the tree in a way similar to the univariate case in Section~\ref{sec:latent_state}. We repeat this procedure for $1000$ MC runs and average the tree-conditional posterior predictive to estimate the unconditional posterior mean. We use this approach to estimate the posterior mean (i.e., the posterior predictive density) under both the partial-likelihood and full-likelihood specifications.

\subsubsection{Results for $L_1, L_{\infty}$}
\begin{figure}
        \hspace{-1cm}
        \includegraphics[width=1.1\textwidth]{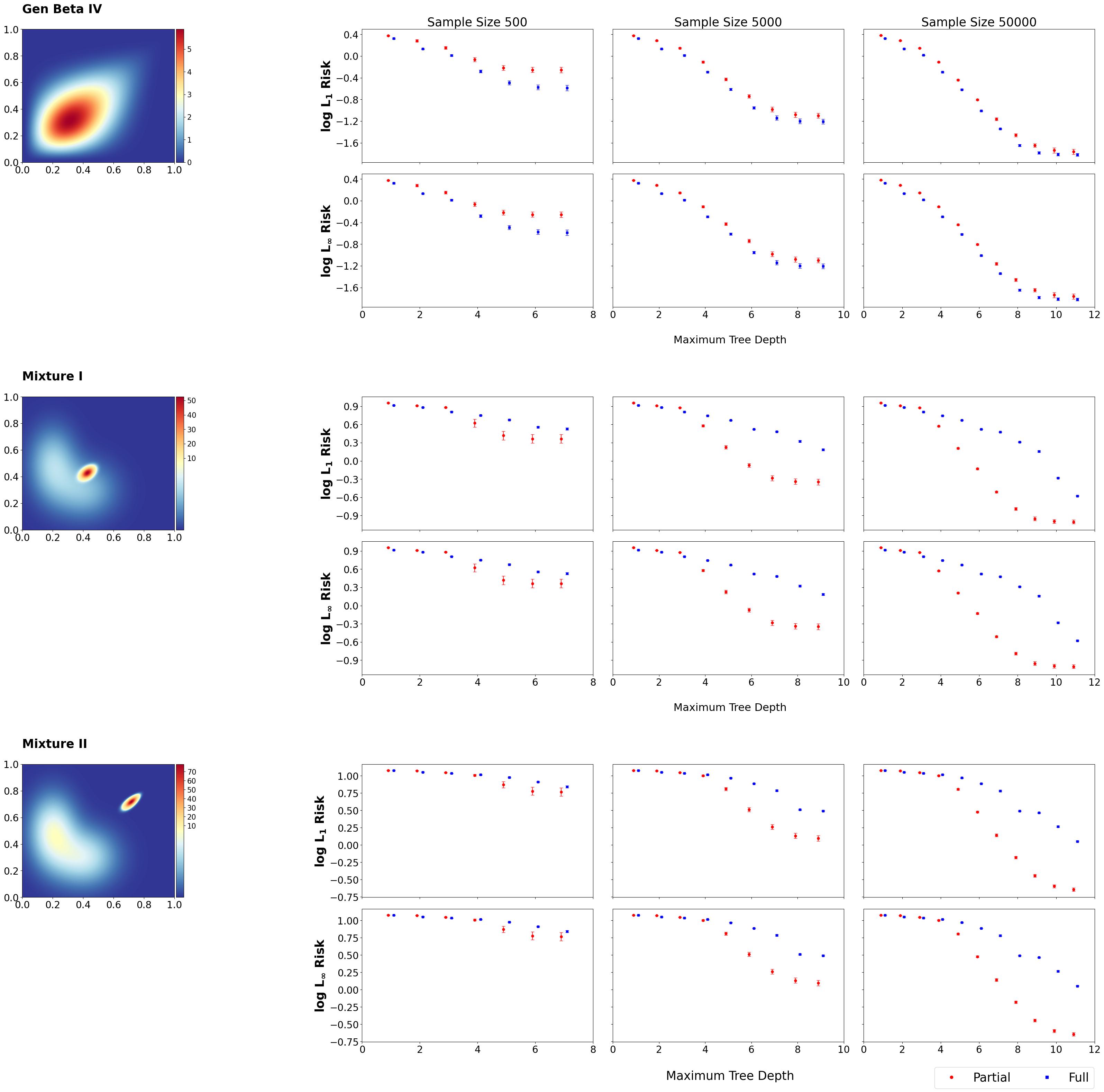}
        \caption{Simulation results for log $L_1, L_{\infty}$ risks in the bivariate examples. }
        \label{fig:2D_L1_LI}
\end{figure}

\ref{fig:2D_L1_LI} reports the log $L_1, L_{\infty}$ risks obtained under the scenarios described in Section~\ref{sec:num_ex_bivariate}. Again, they suggest similar patterns to those already discussed based on the $L_2$ risk, with the partial likelihood tree being the preferable model in the presence of high-density spikes. However, the discrepancy between the two tree specifications is less neat according to these two metrics. For example, the $L_1$ distance for the Mixture examples displays better performance of the partial model only for tree depths in the medium-high range. The $L_1$ partial model gains are less evident for the Generalized Beta II than in the $L_2$ case, and the Generalized Beta III shows comparable (if not better) performance of the full-likelihood tree. On the other hand, $L_{\infty}$ shows major advantages of the partial model in the presence of high spikes and for higher sample sizes ($5000-50000$). However, it displays a rather ``jumpy'' behavior, probably due to the extremely local nature of such a metric. 

\newpage

\newpage 

\begin{figure}
        \hspace{-1cm}
        \includegraphics[width=1.1\textwidth]{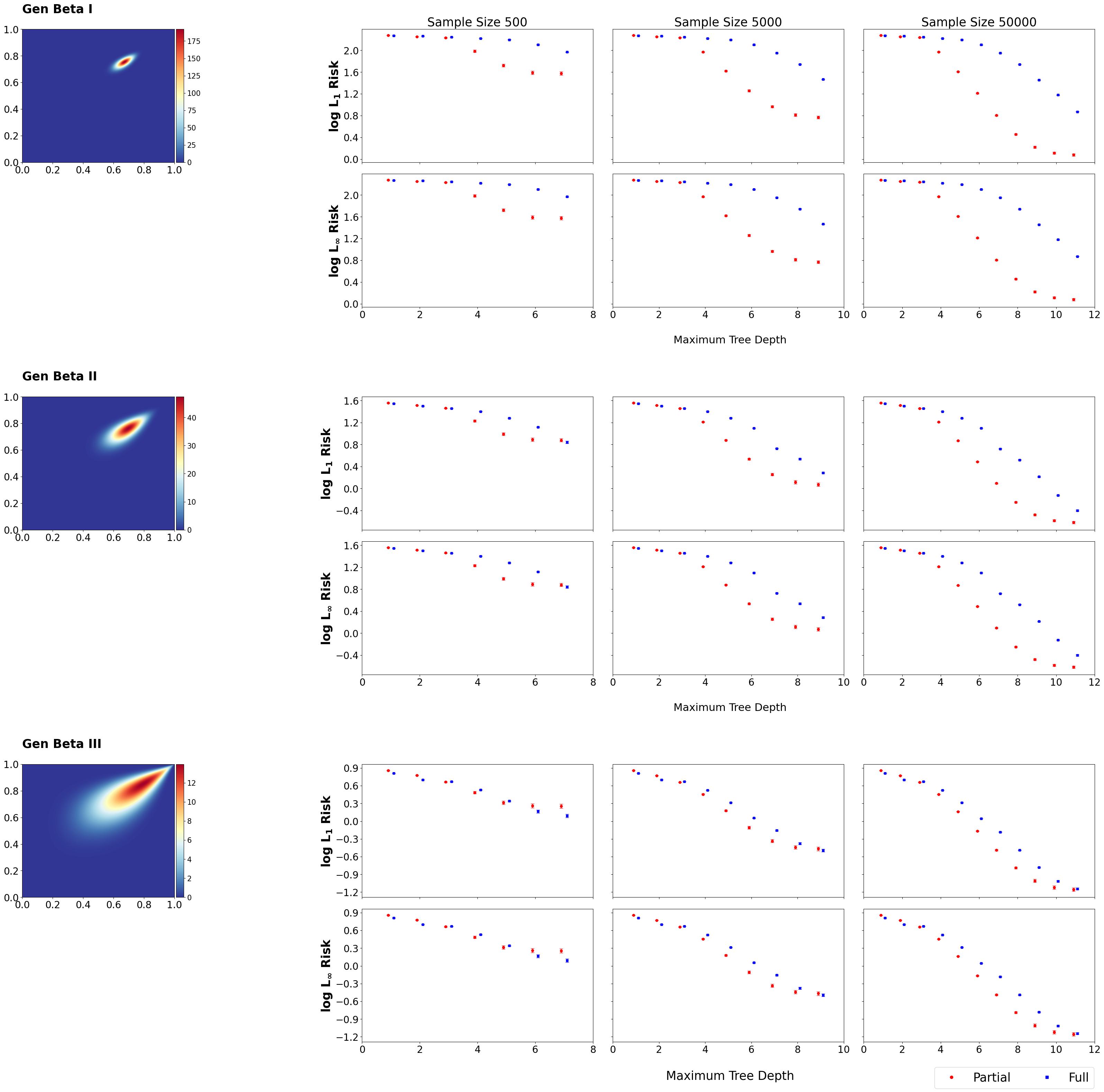}
\end{figure}

\newpage 

\end{document}